\documentclass[useAMS,usenatbib,usegraphicx,longabstract]{mn2e}

\usepackage{times}
\usepackage{epsfig}
\usepackage{epstopdf}
\usepackage{amsmath}

\newcommand{\kms}{$\rm km\,s^{-1}$}

\newcommand{\OI}{[{\sc O$\,$i}]}
\newcommand{\hbo}{$\rm H\beta_o$}
\newcommand{\hb}{$\rm H\beta$}
\newcommand{\mgfep}{$\rm [MgFe]'$}
\newcommand{\mgfe}{$\rm [MgFe]$}
\newcommand{\zh}{$\rm [Z/H]$}

\newcommand{\fem}{$\rm \langle Fe\rangle$}
\newcommand{\gammab}{$\rm \Gamma_b$}
\newcommand{\Dix}{$\rm \Delta_{i,X}$}

\newcommand{\placefigone}{
\begin{figure}
\begin{center}
\includegraphics[width=\columnwidth,bbllx=27pt,bblly=28pt,bburx=424pt,bbury=623pt]{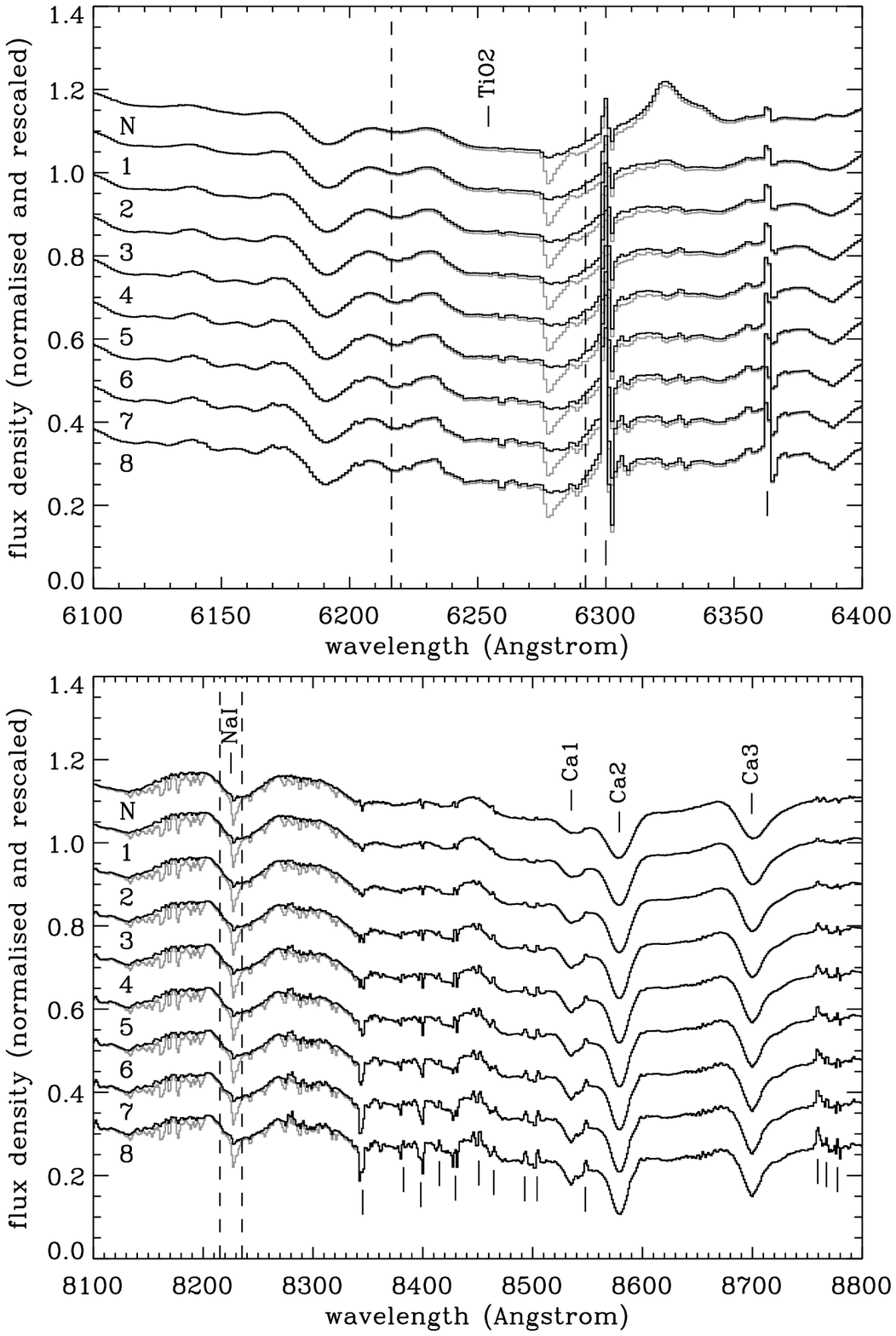}
\end{center}
\caption[]{Quality of our correction for telluric atmospheric
  absorption, as shown by applying the {\sc molecfit} procedure to the
  MUSE aperture spectra extracted from the annular regions shown in
  Fig.~\ref{fig:two}. These spectra have been normalised and offset
  for convenience of display, and arranged from top to bottom as a
  function of distance from the center of M87. Uncorrected spectra are
  shown by the grey lines. The top panel shows the wavelength region
  near the observed position of the TiO2 absorption feature, whereas
  the lower panel cover the region occupied by both the NaI absorption
  and Ca triplet. The vertical lines bracket the central passbands
  defining the IMF-sensitive TiO2 and NaI indices (see
  Tab.~\ref{tab:two}). Spectral regions most affected by sky emission
  are indicated by the vertical lines plotted below the outer aperture
  spectrum.} \label{fig:one}
\end{figure}
}

\newcommand{\placefigtwo}{
\begin{figure}
\begin{center}
\includegraphics[width=\columnwidth]{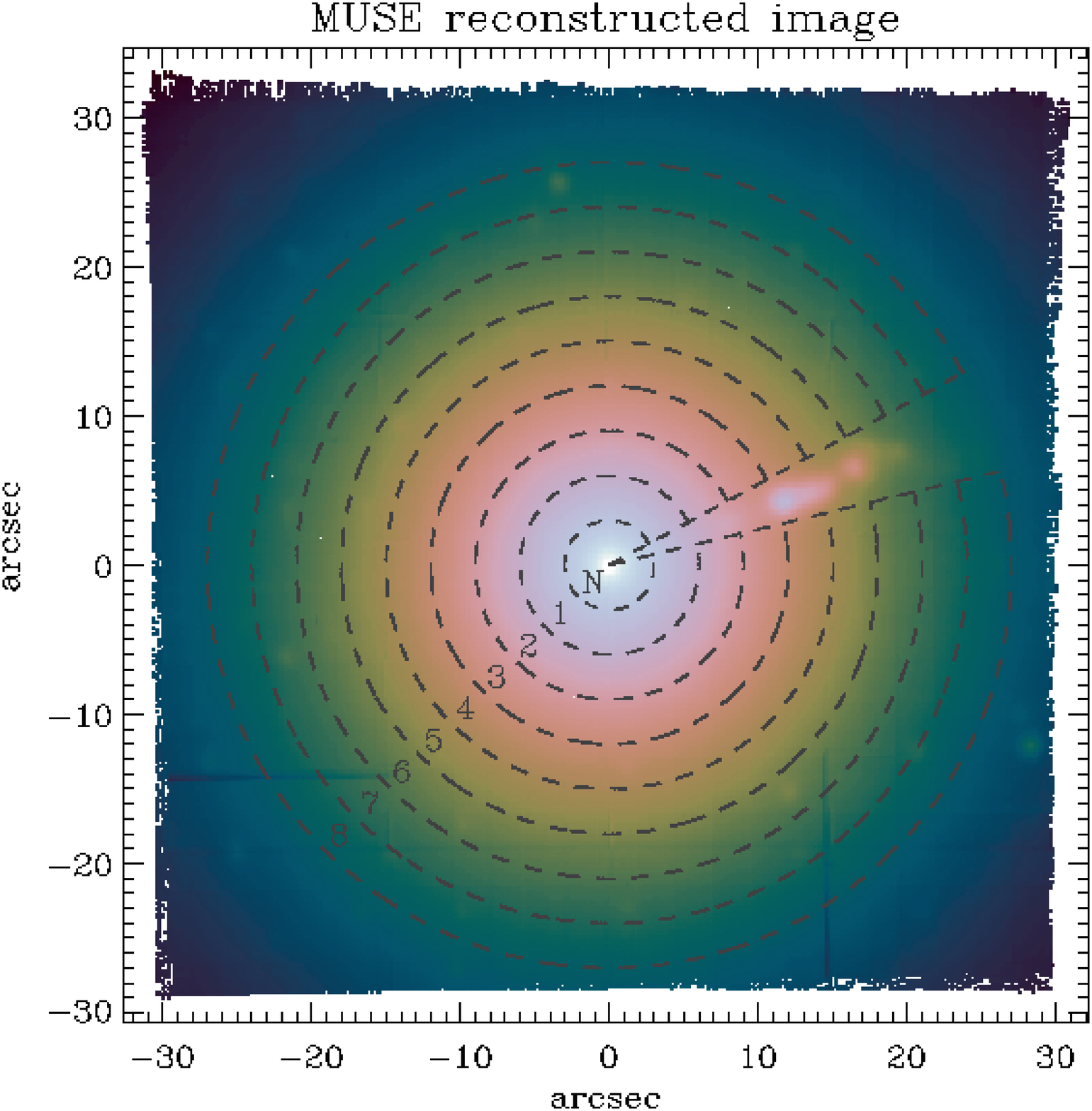}
\end{center}
\caption[]{MUSE reconstructed image for the central regions of M87,
  showing the eight, nearly circular annular regions over which we
  extracted the aperture spectra shown in Figs.~\ref{fig:one} and
  \ref{fig:three}. Regions contaminated by synchrotron emission
  associated to the jet of M87 have been excluded in these
  apertures. A nuclear spectrum has also been extracted but will only
  serve for illustrative purposes since such a spectrum is heavily
  contaminated by non-thermal continuum emission.} \label{fig:two}
\end{figure}
}

\newcommand{\placefigthree}{
\begin{figure}
\begin{center}
\includegraphics[width=\columnwidth,bbllx=53pt,bblly=28pt,bburx=424pt,bbury=623pt]{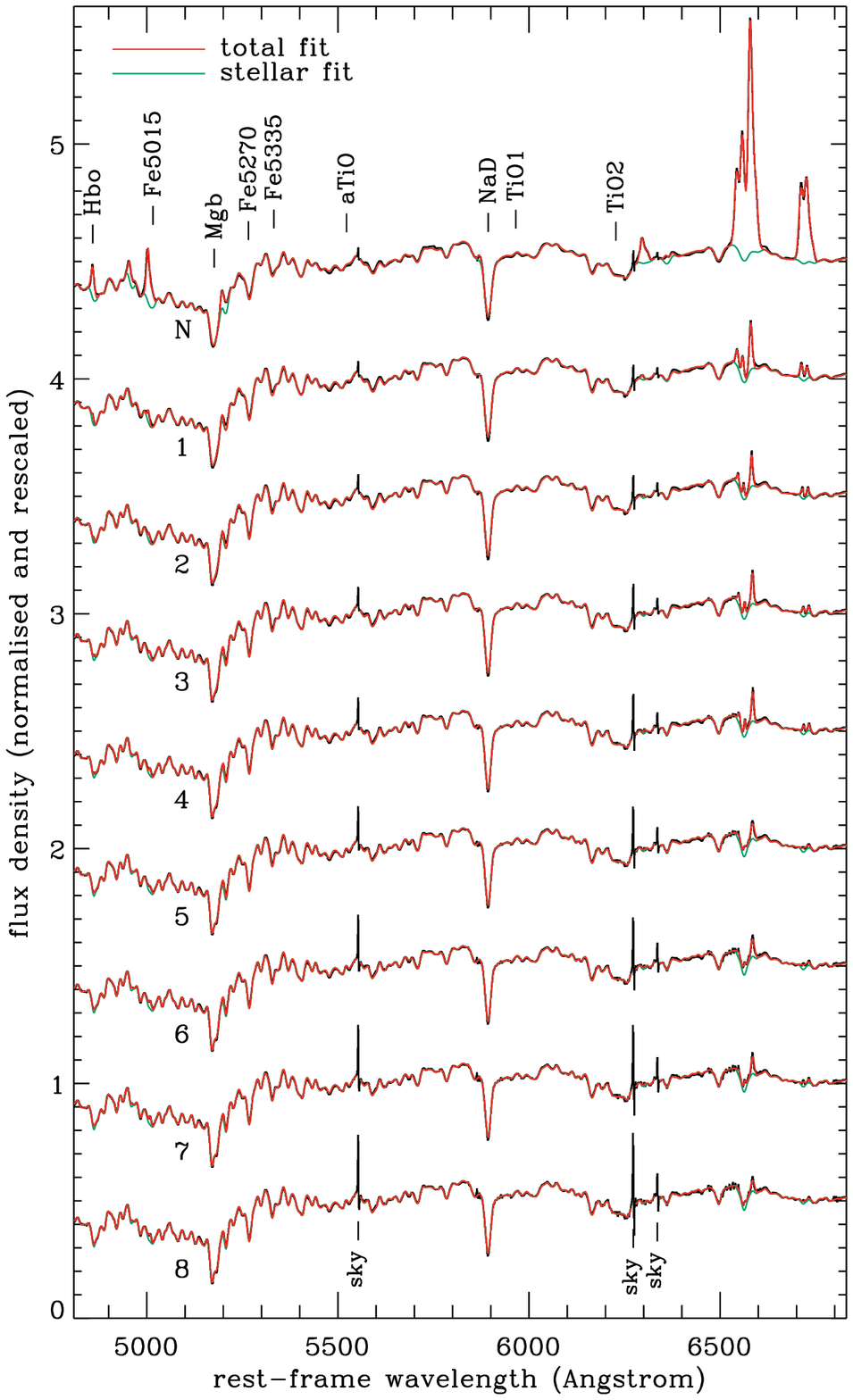}
\end{center}
\caption[]{MUSE aperture spectra extracted from the annular regions
  shown in Fig.~\ref{fig:two} over the wavelength range of our {\sc
    GandALF\/} fit (red and green lines for total and stellar fit
  only, respectively). These spectra have been normalised and offset
  for convenience of display, and arranged from top to bottom as a
  function of distance from the center. The position of the adopted
  IMF-sensitive absorption features (see Tab.~\ref{tab:two}) that fall
  in this wavelength range is also shown, together with that of the
  NaD doublet that responds to the [Na/Fe] abundance and of the
  classical Hb, Mgb, Fe5270 and Fe5335 Lick indices that will serve to
  constrain the stellar age, metallicity and the abundance of
  $\alpha$-elements. Regions affected by sky emission (at an observed
  wavelength of 5577, 6300 and 6330 \AA) have been omitted in the
  fit.} \label{fig:three}
\end{figure}
}

\newcommand{\placefigfour}{
\begin{figure*}
\begin{center}
\includegraphics[width=\textwidth]{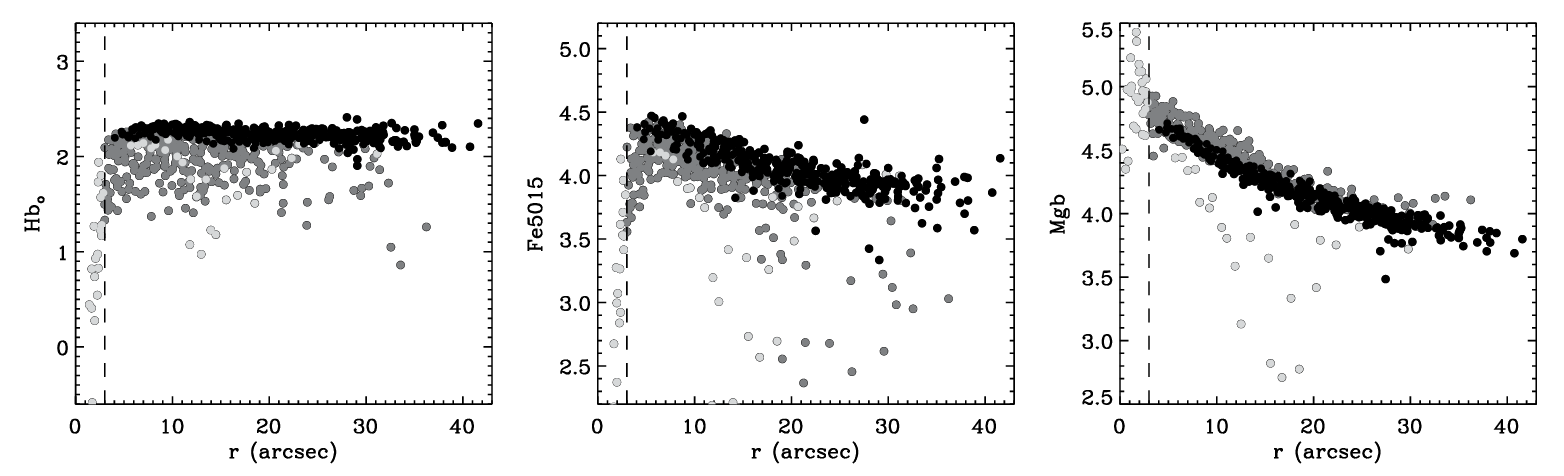}
\end{center}
\caption[]{Radial profiles for the strength of the \hbo, Fe5015 and
  Mgb Lick absorption-line indices that enter our stellar age,
  metallicity and $\alpha$-elements abundance measurements, as
  measured in our Voronoi-binned spectra
  (\S\ref{subsec:SN300_bins}). The dark grey points in each panel show
  Voronoi bins that will be excluded from our analysis due to the
  adverse impact of nebular emission, whereas the light grey points
  indicate bins discarded due to the presence of a non-thermal
  continuum associated either to the jet or AGN of M87. Such a
  featureless continuum tends to dilute absorption lines and decrease
  the value of their correponding line-strength indices, similar to
  the case of line infill from H$\beta$ and [{\sc O$\,$iii}] in the
  case of the \hbo\ and Fe5015 indices, respectively. [{\sc N$\,$i}]
  emission, on the other hand, falls on the red continnum passband of
  the Mgb index and thus leads to an artificial increase of its
  value. The index values shown here were computed after bringing the
  Voronoi-binned spectra to a common kinematic broadening
  corresponding to a stellar velocity dispersion of 360 \kms, in order
  to allow a direct comparison at different galactic
  radii.} \label{fig:four}
\end{figure*}
}

\newcommand{\placefigfive}{
\begin{figure*}
\begin{center}
\includegraphics[width=\textwidth]{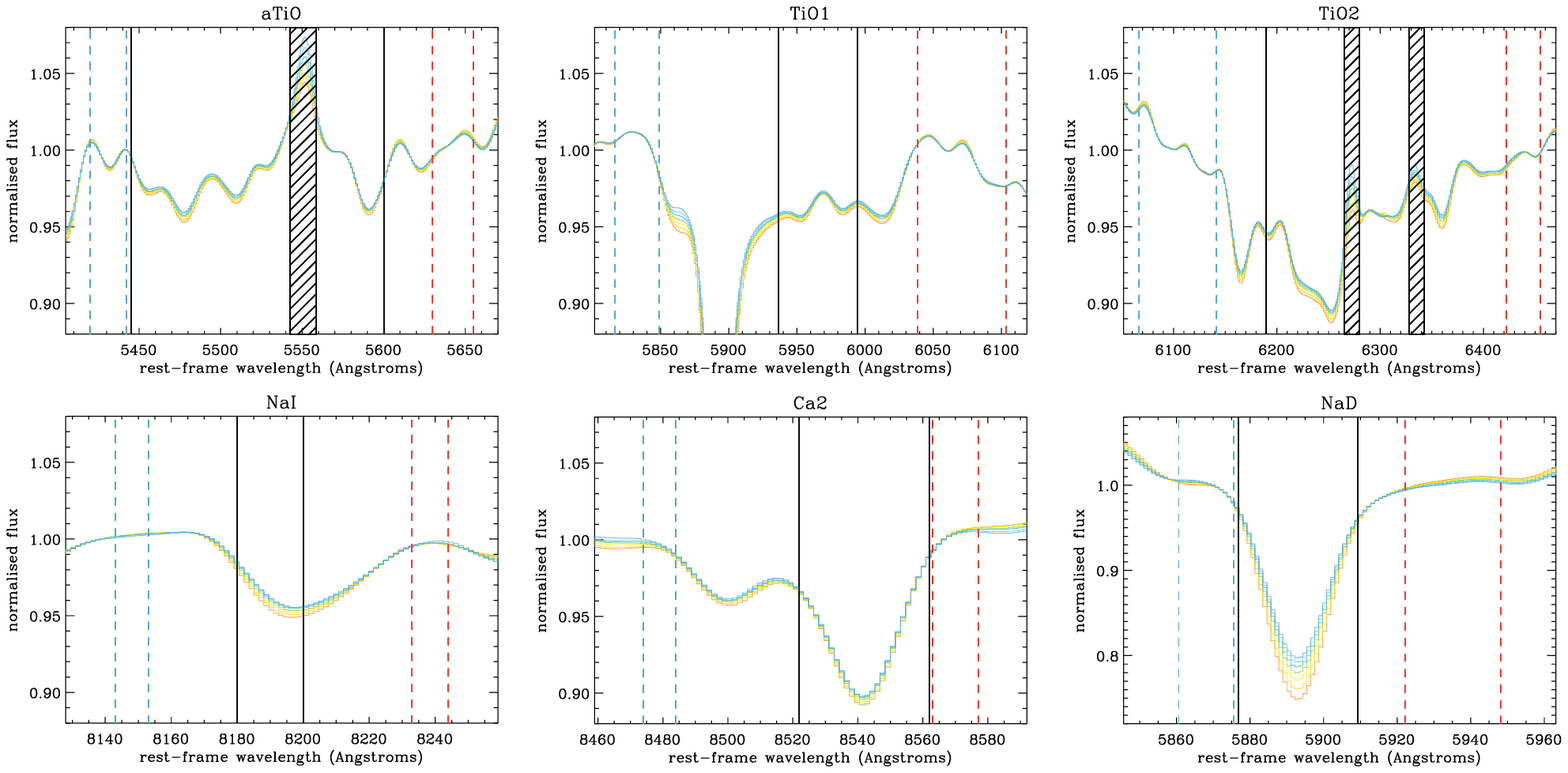}
\end{center}
\caption[]{Detailed view of the annular aperture spectra of
  Fig.~\ref{fig:three} (excluding the nuclear spectrum) illustrating the
  radial variation for the strength for the IMF-sensitive
  absorption-line features aTiO, TiO1, TiO2, NaI and Ca2, as well as
  for the NaD absorption on which our [Na/Fe] estimates will be
  based. In each panel, the aperture spectra are colour-coded
  according to the radial distance they probe and have been brought to
  a common kinematic broadening corresponding to a stellar velocity
  dispersion of 360 \kms. Vertical solid lines indicate the index
  bandpass, whereas the blue and red dashed lines the blue and red
  continuum regions, respectively (see Tab.~\ref{tab:two}). All
  spectra are normalised to the pseudo-continuum level in the index
  bandpass. Vertical hatched boxes shows regions affected by sky
  emission. The Ca1 and Ca3 features show similarly clear variations
  to Ca2 and are omitted here for clarity.} \label{fig:five}
\end{figure*}
}

\newcommand{\placefigsix}{
\begin{figure*}
\begin{center}
\includegraphics[width=\textwidth]{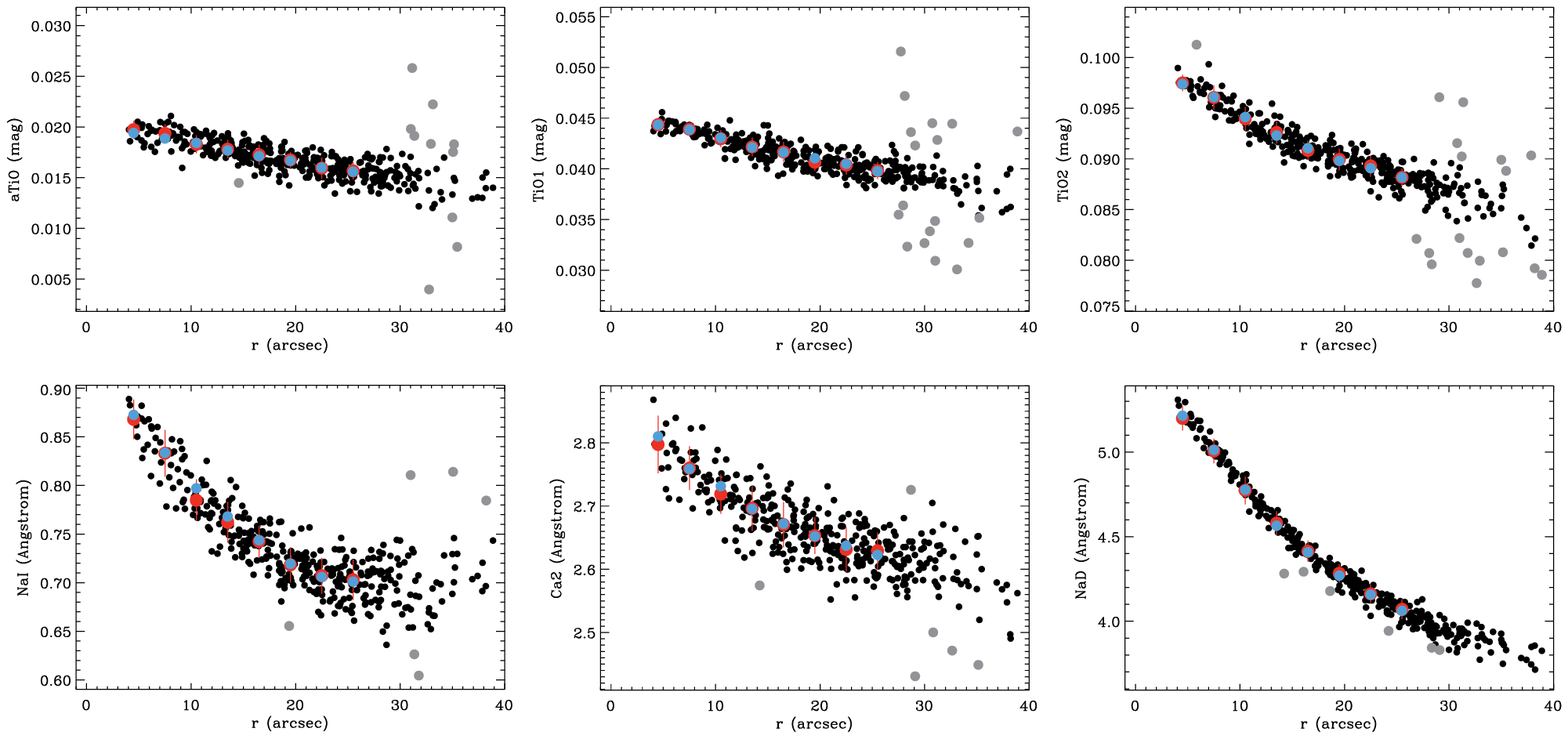}
\end{center}
\caption[]{Radial trend for the values of the IMF-sensitive aTiO,
  TiO1, TiO2, NaI and Ca2 indices and for the NaD index, as measured
  in our annular aperture spectra (blue bullets) or in Voronoi-binned
  spectra with formal $S/N$ of 300 (black \& grey small bullets). The
  red bullets with errors bars show the median values of the latter
  measurements and their scatter, which provides a conservative
  estimate of the uncertainties in our index measurements given the
  circular symmetry of M87. 
  The index values shown here were computed after bringing the spectra
  to a common kinematic broadening corresponding to a stellar velocity
  dispersion of 360 \kms, in order to allow a direct comparison for
  strength of absorption lines at different galactic radii (for our
  analysis indices are measured on the original spectra, see
  \S~\ref{subsec:LS_measurements}).
  Voronoi bins significantly affected by the non-thermal continuum
  associated to the jet and AGN of M87, or by nebular emission, are
  not shown and were not used to compute the mean and standard
  deviations on the Voronoi-bin measurements.
  The grey points show 3$\sigma$ outliers from a 3rd-order polynomial
  fit to the radial gradients of all line-strength indices used in
  this work, which are also excluded during our
  analysis.} \label{fig:six}
\end{figure*}
}

\newcommand{\placefigseven}{
\begin{figure}
\begin{center}
\includegraphics[width=0.9\columnwidth,bbllx=55pt,bblly=35pt,bburx=406pt,bbury=342pt]{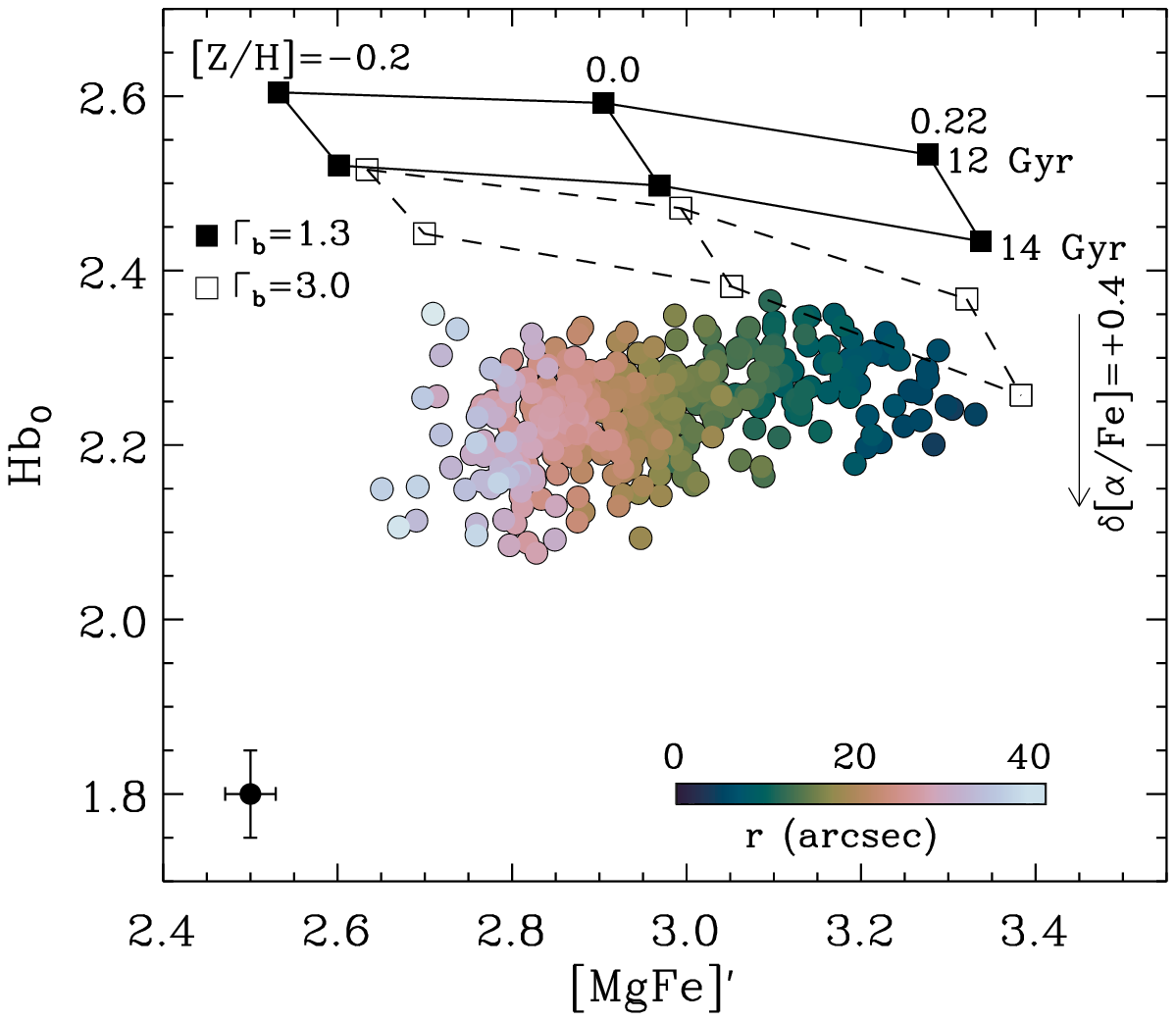}
\end{center}
\caption[]{\hbo\--\mgfep\ index-index diagram as measured in our final
  sample of Voronoi-binned spectra, that is, by excluding regions
  affected by nebular emission, a non-thermal continuum associated the
  AGN and/or jet of M87 (see Fig.~\ref{fig:four}) or where
  systematic effects led to unreliable line-strength measurements (see
  Fig.~.\ref{fig:six}).
  Shown are also MIUSCAT model grids for varying stellar age and
  metallicity, drawn with solid and dashed lines for a Kroupa IMF
  (corresponding to a ${\rm \Gamma_b = 1.3}$, see \S~4.1) and for a
  bottom-heavy IMF (${\rm \Gamma_b = 3.0}$), respectively. The impact
  of an ehnanced $\alpha$-element is also shown by the
  downward-pointing vertical arrow \citep[from the models
    of][]{Vaz15}. The data points are color-coded according to
  distance from the center, which highlights the presence of a
  well-established negative metallicity gradient in ETGs. Typical
  errors are also shown at the lower left corner of this
  figure.} \label{fig:seven}.
\end{figure}
}

\newcommand{\placefigeight}{
\begin{figure*}
\begin{center}
\includegraphics[width=\textwidth]{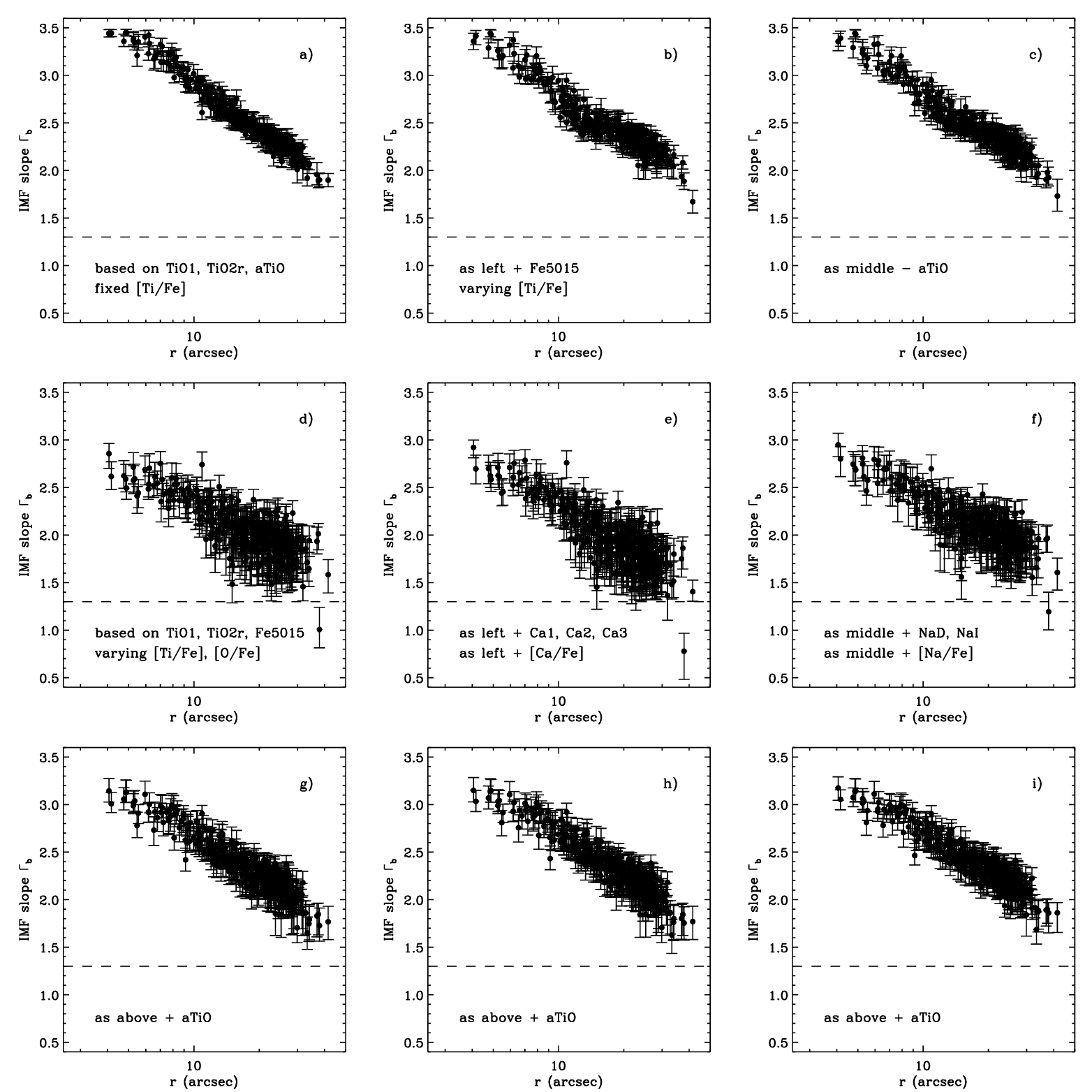}
\end{center}
\caption[]{Radial profiles for the best-fitting slope \gammab\ for a
  low-mass tapered ``bimodal'' IMF, obtained by mimimising the
  expression given by Eq.~1 while using different sets of
  absorption-line indices and by either holding or varying the
  abundance of different elements. Irrespective of the latter choices
  the IMF slope \gammab\ is always found to decrease with radius and
  to display values well above ${\rm \Gamma_b = 1.3}$ as in the case
  of a Milky-Way like, Kroupa IMF (horizontal
  lines).} \label{fig:eight}
\end{figure*}
}

\newcommand{\placefignine}{
\begin{figure*}
\begin{center}
\includegraphics[width=\textwidth]{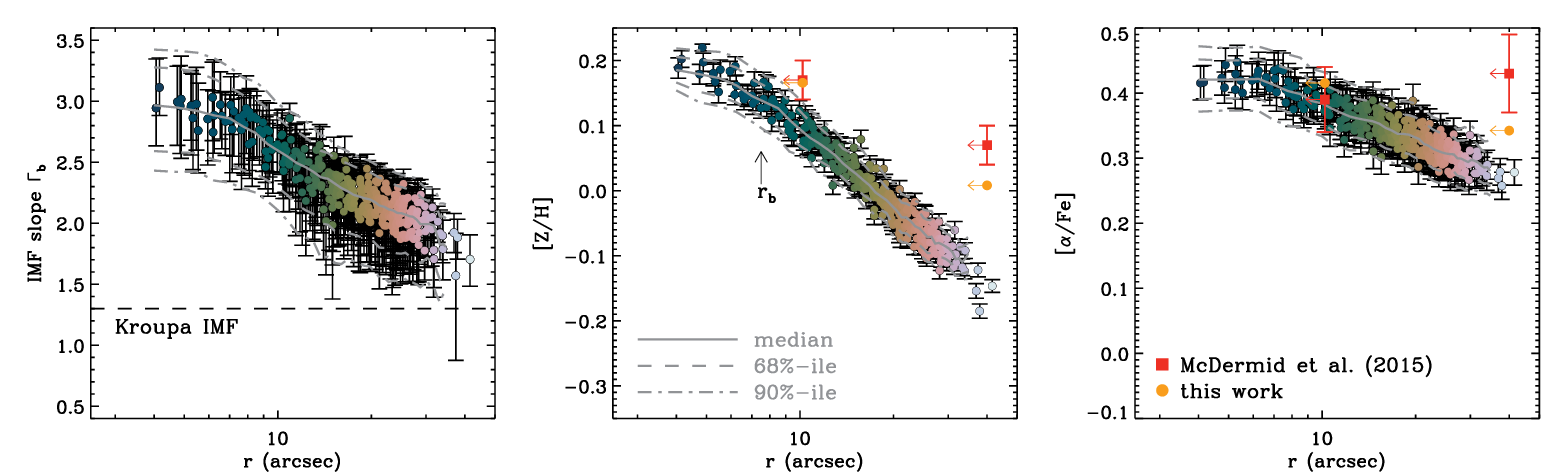}
\end{center}
\caption[]{Final radial profiles for the the best-fitting slope
  \gammab\ for a low-mass tapered ``bimodal'' IMF (left panel), as
  obtained from combining the results based on different combinations
  of absorption-line indices and while either holding or varying the
  abundance of various elements (see Fig.~\ref{fig:eight}), as well as
  for the stellar metallicity [Z/H] and the $\alpha$-elements
  abundance [$\alpha$/Fe] (middle and right panels, respectively). In
  each panel the grey solid line shows median values within
  3\arcsec-wide radial bins, whereas the grey dashed and dot-dashed
  lines show the 68\% and 90\% confidence levels around such a median,
  respectively. Points are color-coded according to distance from the
  center of M87 as in Fig.~\ref{fig:seven} and the horizontal dashed
  line in the left panel shows the \gammab\ value for a Kroupa IMF, as
  in Fig.~\ref{fig:eight}. In the middle and right panels, the two
  orange circles and red squares, each with a left-ward pointing
  arrow, indicate the luminosity-weighted [Z/H] and [$\alpha$/Fe]
  values inside $R_e/8$ and $R_e/2$, as computed here (after
  extrapolating the gradients up to $r=2\arcsec$) and in
  \citet[][based on SAURON data]{McD15}, respectively. The vertical
  arrow in the middle panel indicate the break radius $r_b$ where the
  surface brightness flattens and departs from a single Sersic
  profile (see Fig.~4 of \citealp{Cot06}).} \label{fig:nine}
\end{figure*}
}

\newcommand{\placefigten}{
\begin{figure}
\begin{center}
\includegraphics[width=\columnwidth]{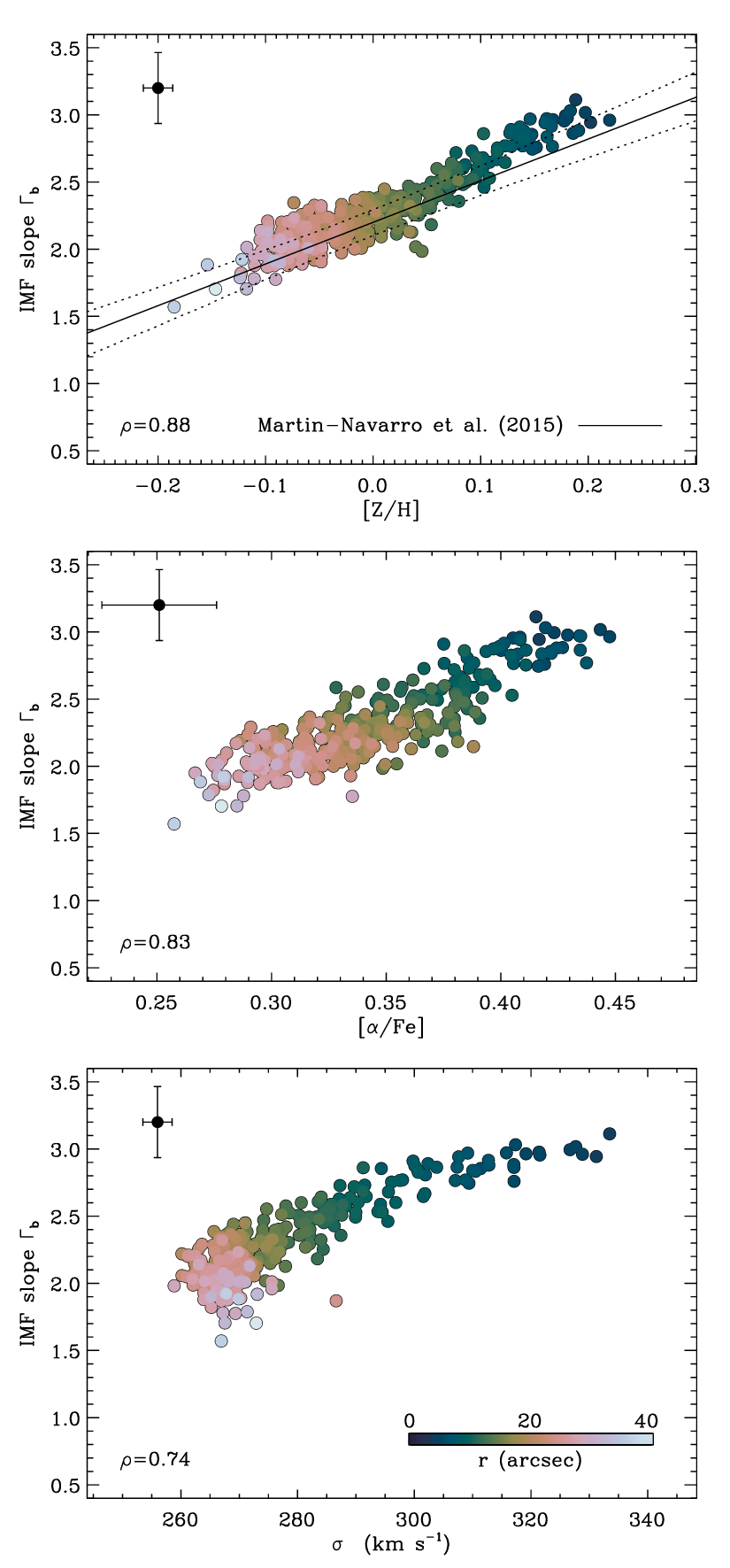}
\end{center}
\caption[]{Correlations between the inferred slope \gammab\ for a
  low-mass tapered ``bimodal'' IMF and our measurements for the
  stellar metallicity [Z/H] (top panel), $\alpha$-elements abundance
  [$\alpha$/Fe] (midlle panel) and stellar velocity dispersion
  $\sigma$ (lower panel). In each panel the typical errors in both
  plotted quantities are shown in the top left corner, whereas the
  value of the Spearman's rank correlation coefficient $\rho$ is
  reported in the lower left corner.  Data points are color-coded
  according to distance from the center as in Fig.~\ref{fig:seven}.
  The top panel also shows IMF--metallicity relation of
  \citet{Mar15b}, which our MUSE data seem to follow rather well. Even
  though such an agreement supports the idea that the IMF shape is
  more tighly connected to the stellar metallicity then to other
  stellar population parameters, in the case of M87 the
  \gammab\ values would appear to follow fairly well also the
                  [$\alpha$/Fe] values (middle panel). On the other
                  hand, our analysis confirms that $\sigma$ is not a
                  particularly good tracer for the radial variation of
                  the IMF in early-type.  galaxies.} \label{fig:ten}
\end{figure}
}

\newcommand{\placefigeleven}{
\begin{figure}
\begin{center}
\includegraphics[width=\columnwidth]{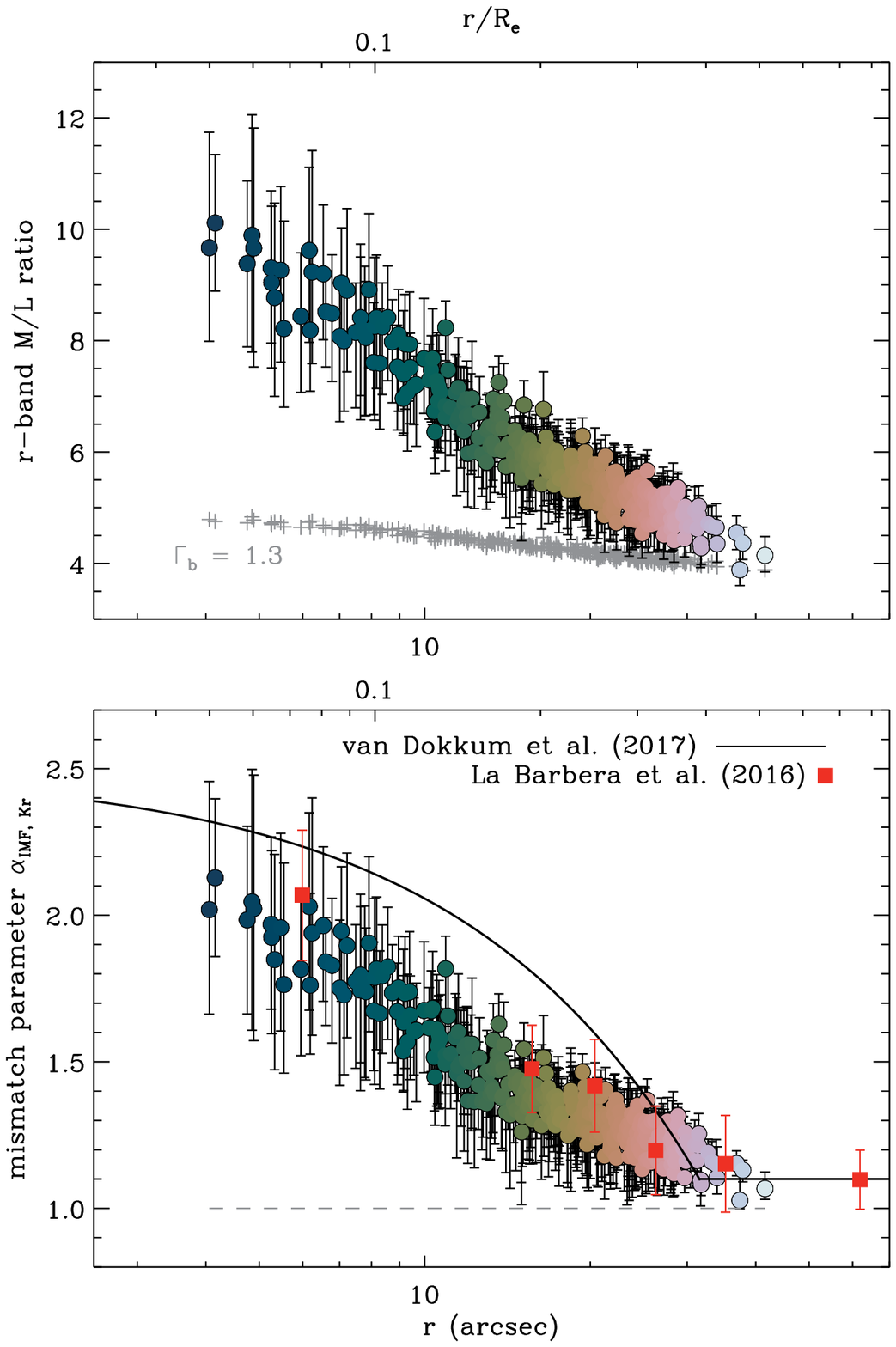}
\end{center}
\caption[]{Radial profile for the r-band stellar mass-to-light ratio
  (top panel) and mismatch parameter $\alpha_{\rm IMF}$ (lower panel)
  corresponding to our best-fitting slope \gammab\ of a low-mass
  tapered ``bimodal'' IMF, as well as to our best values for the
  stellar metallicity and [$\alpha$/Fe] ratio. The grey points in top
  panel show the $(M/L)_{\rm Kr}$ values corresponding to a Kroupa IMF
  (when $\Gamma_b = 1.3$ ), which also serves to compute the mismatch
  parameter $\alpha_{\rm IMF, Kr} = (M/L)/(M/L)_{\rm Kr}$ that is
  shown in the lower panel where the grey horizontal line indicates
  the case of a Kroupa IMF. In the lower panel our $\alpha_{\rm IMF,
    Kr}$ profile is also compared to the mean trend recently inferred
  by \citet{vDo17} and to the $\alpha_{\rm IMF, Kr}$ gradient measured
  by \citet{LaB16} in the early-type galaxy XSG1. To this effect at
  the top of each panel we also show the radial scale in units of the
  effective radius $R_e$.} \label{fig:eleven}
\end{figure}
}

\newcommand{\placefigtwelve}{
\begin{figure}
\begin{center}
\includegraphics[width=\columnwidth]{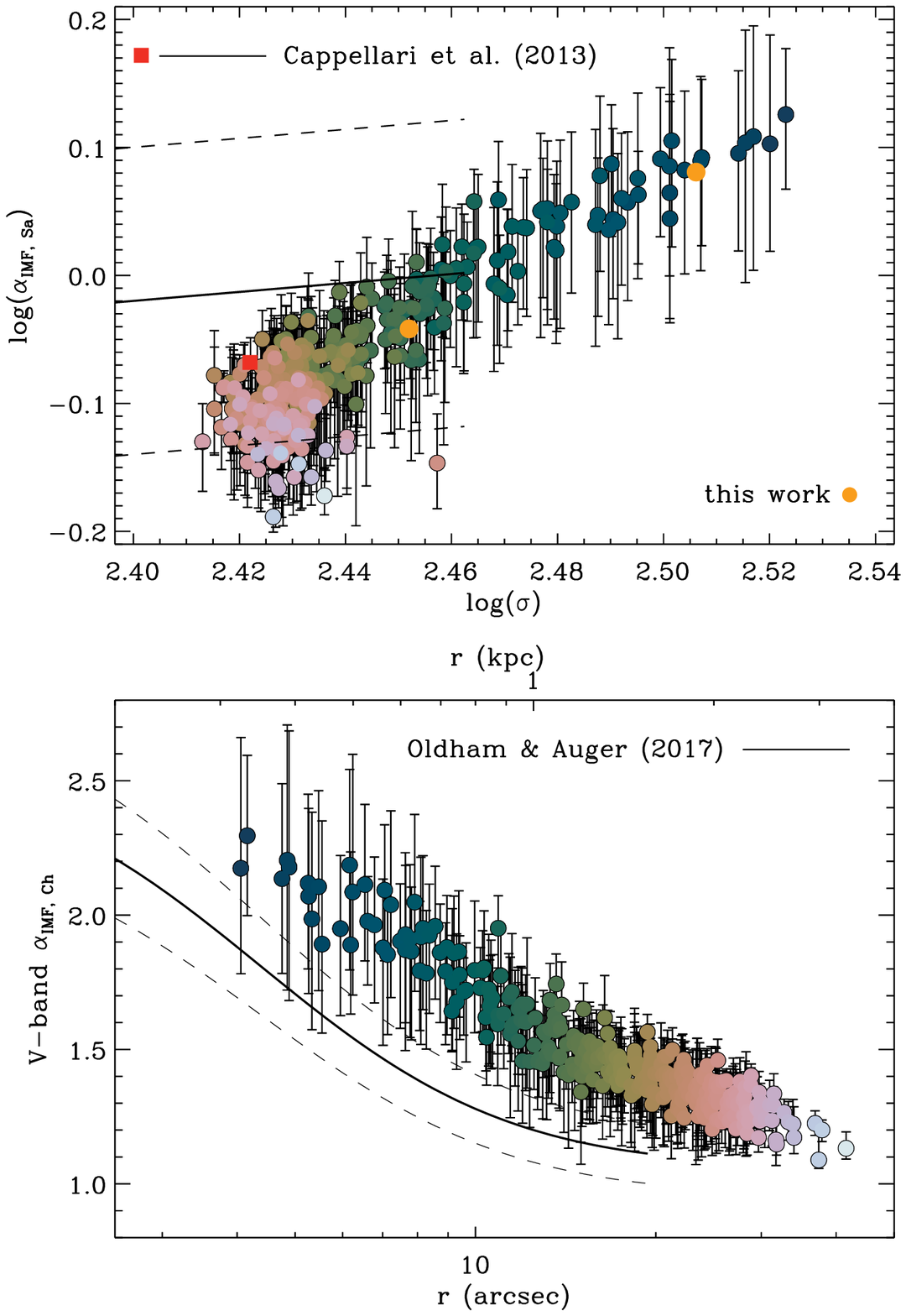}
\end{center}
\caption[]{Top panel: r-band mismatch parameter rescaled with respect
  to a Salpeter IMF and plotted against stellar velocity dispersion
  $\sigma$ in order to be compared to the global value for
  $\alpha_{\rm IMF, Sa}$ in M87 (red square) and across different
  ETGs , (solid and dashed lines) as function of global stellar
  velocity dispersion (within 1 $R_e$) as derived through dynamical
  and stellar population modelling over the course of the ATLAS$^{\rm
    3D}$ survey (Cappellari et al. 2013). As in Fig.~\ref{fig:nine}
  the orange circles correspond to our luminosity-weighted integrated
  measurement inside $R_e/8$ and $R_e/2$.
  Lower panel: radial profile for the V-band mismatch parameter with
  respect to a Chabrier IMF $\alpha_{\rm IMF, Ch}$ compared to the
  predictions for this same quantity by Oldham \& Auger (2017), who
  allow for a varying $M/L$ ratio in their dynamical models. Although
  systematically below our values, the $\alpha_{\rm IMF, Ch}$
  gradient of Oldham \& Auger parallels remarkably well our
  $\alpha_{\rm IMF, Ch}$ profile.} \label{fig:twelve}
\end{figure}
}

\newcommand{\placefigNaD}{
\begin{figure}
\begin{center}
\includegraphics[width=\columnwidth,bbllx=28pt,bblly=33pt,bburx=424pt,bbury=321pt]{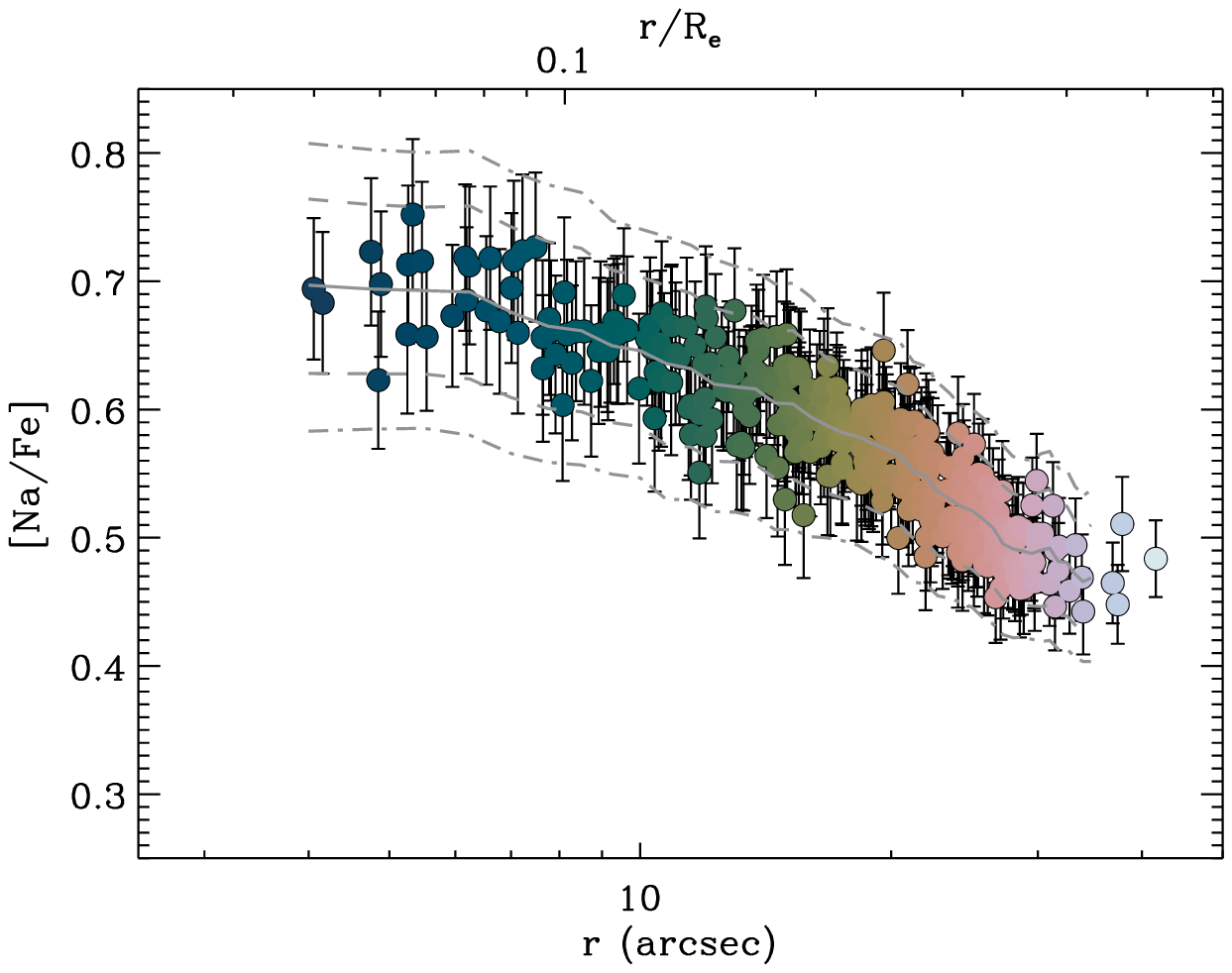}
\end{center}
\caption[]{Radial profile for the [Na/Fe] abundance in M87, as derived
  in our Voronoi-binned spectra. As in Fig.~\ref{fig:nine} the grey
  solid line shows median values within 3\arcsec-wide radial bins,
  whereas the grey dashed and dot-dashed lines show the 68\% and 90\%
  confidence levels around such a median. Data points are also
  color-coded according to distance from the center as in
  Fig.~\ref{fig:seven}.} \label{fig:NaD}
\end{figure}
}

\newcommand{\placetabone}{
\begin{table}
\caption{Aperture spectra basic properties}
\label{tab:one}
\begin{center}
\begin{tabular}{rrrrrr}
\hline
ID & radius  & N$_{\rm spax}$ & $S/N$ & $S/rN$ & $\sigma_{\star}$ \\
(1)& (2)     & (3)            & (4) & (5)  & (6)              \\
\hline
 N & $<3$    &   675         & 1465 & 174  & 386.2            \\ 
 1 & 3--6    &  2033         & 2294 & 195  & 331.3            \\
 2 & 6--9    &  3379         & 2625 & 201  & 305.9            \\
 3 & 9--12   &  4739         & 2774 & 212  & 289.4            \\
 4 & 12--15  &  6094         & 2793 & 218  & 278.9            \\
 5 & 15--18  &  7470         & 2768 & 229  & 271.4            \\
 6 & 18--21  &  8792         & 2703 & 229  & 267.1            \\
 7 & 21--24  & 10164         & 2649 & 229  & 265.4            \\
 8 & 24--27  & 11500         & 2543 & 235  & 265.2            \\
\hline
\end{tabular}
\end{center}
\begin{minipage}{\columnwidth}{\it Notes\/}: (1)~Aperture number. 
  (2)~Radial range, in arcseconds. (3)~Number of original
  $0\farcs2\times0\farcs2$ MUSE spaxels within the
  aperture. (4)~Formal signal-to-noise ratio (at 5100\AA), as expected
  from propagation of the statistical uncertainties returned by the
  data reduction. (5)~Observed signal-to-residual-noise value reached
  by our best {\sc GandALF\/} fits (over the entire fitted
  range). (6)~Stellar velocity dispersion, in \kms, as returned by our
  {\sc pPXF\/} fit.
\end{minipage}
\end{table}
}    

\newcommand{\placetabtwo}{
\begin{table}
\caption{Definition of the most IMF-sensitive indices adopted in this study}
\label{tab:two}
\begin{center}
\begin{tabular}{lcccl}
\hline
Index & $\lambda_{bb}$ -- $\lambda_{br}$ & $\lambda_{cb}$ -- $\lambda_{cr}$ & $\lambda_{rb}$ -- $\lambda_{rr}$ & ref. \\
(1)   & (2)      & (3)      & (4) & (5) \\
\hline
 aTiO & 5420.0 -- 5442.0 & 5445.0 -- 5600.0 & 5630.0 -- 5655.0 & 1 \\
 TiO1 & 5816.6 -- 5849.1 & 5936.6 -- 5994.1 & 6038.6 -- 6103.6 & 2 \\
 TiO2 & 6066.6 -- 6141.6 & 6189.6 -- 6265.0 & 6422.0 -- 6455.0 & 3 \\
 NaI  & 8143.0 -- 8153.0 & 8180.0 -- 8200.0 & 8233.0 -- 8244.0 & 3 \\
 Ca1  & 8474.0 -- 8484.0 & 8484.0 -- 8513.0 & 8563.0 -- 8577.0 & 4 \\
 Ca2  & 8474.0 -- 8484.0 & 8522.0 -- 8562.0 & 8563.0 -- 8577.0 & 4 \\
 Ca2  & 8619.0 -- 8642.0 & 8642.0 -- 8682.0 & 8700.0 -- 8725.0 & 4 \\
\hline
\end{tabular}
\end{center}
\begin{minipage}{\columnwidth}{\it Notes\/}: (1)~Index name. 
(2)~Blue and red wavelength limit for the blue pseudo-continuum
  passband.  (3)~Blue and red wavelength limit for the index passband.
  (4)~Blue and red wavelength limit for the red pseudo-continuum
  passband.  (5)~Source for the index definition: 1 -- \citet{Spi14},
  2 -- \citet{Tra98}, 3 -- \citet{LaB15}, 4 -- \citet{Cen01}.
\end{minipage}
\end{table}
}

\title[IMF and Na/Fe gradients in NGC~4486]{MUSE observations of M87:
  radial gradients for the stellar initial-mass function and the
  abundance of Sodium}

\author[Sarzi et al.]{Marc Sarzi$^{1}$\thanks{E-mail :sarzi@star.herts.ac.uk}\\
  $^{1}$Centre for Astrophysics Research, University of Hertfordshire, 
  College Lane, Hatfield, Herts, AL10 9AB, UK\\}

\author[Sarzi et al.] {\parbox{\textwidth}{Marc
    Sarzi,$^{1}$\thanks{E-mail:\texttt{ m.sarzi@herts.ac.uk}}
 Chiara Spiniello,$^{2}$ Francesco La Barbera,$^{2}$ Davor Krajnovi\'c,$^{3}$ Remco van den Bosch
}\vspace{0.4cm}\\
\parbox{\textwidth}{
$^{1}$Centre for Astrophysics Research, University of Hertfordshire,
  Hatfield, Herts AL1 9AB, UK\\
$^{2}$ INAF-Osservatorio Astronomico di Capodimonte, Salita Moiariello, 16, 80131 Napoli, Italy\\
$^{3}$ Leibniz-Institute fur Astrophysics Potsdam (AIP), An der
  Sternwarte 16, D-14482 Potsdam, Germany}
}

\begin{document}
\pagerange{\pageref{firstpage}--\pageref{lastpage}} \pubyear{2017}

\maketitle
\label{firstpage}


%
\begin{abstract}

Based on MUSE integral-field data we present evidence for a radial
variation at the low-mass end of the stellar initial mass function
(IMF) in the central regions of the giant, central-cluster early-type
galaxy NGC~4486 (M87).
We used state-of-the-art stellar population models and the observed
strength of various IMF-sensitive absorption-line features in order to
solve for the best low-mass tapered ``bimodal'' form of the IMF, while
accounting also for radial variations in stellar metallicity, the
overall $\alpha$-elements abundance, and the abundance of individual
elements such as Ti, O, Na and Ca.
Our analysis reveals a strong negative IMF gradient in M87,
corresponding to an exceeding fraction of low-mass stars compared to
the case of the Milky Way toward the innermost regions of M87 that
drops to nearly Milky-way levels by 0.4 $R_e$.
The observed IMF gradient is found to correlate well with both the
radial profile for stellar metallicity and for $\alpha$-elements
abundance but not, unlike the case of global IMF measurements, with
stellar velocity dispersion.
Such IMF variations correspond to over a factor two increase in
stellar mass-to-light $M/L$ ratio compared to the case of a Milky-way
like IMF, consistent with other investigations into IMF gradients in
early-type galaxies (ETGs), including recent dynamical constraints on
$M/L$ radial variations in M87 from dynamical models.
In addition to constraining the IMF in M87 we also looked into the
abundance of Sodium, which turned up to be super-Solar over the entire
radial range of our MUSE observations (with [Na/Fe]$\sim$0.7~dex in
the innermost regions) and to exhibit a considerable negative
gradient. Besides reiterating the importance of constraining the
abundance of Sodium for the purpose of using optical and near-IR
IMF-sensitive Na features, these findings also suggest an additional
role of metallicity in boosting the Na-yields in the central,
metal-rich regions of M87 during its early and brief star formation
history.
Our work adds the case of M87 to the relatively few objects that as of
today have radial constraints on their IMF or [Na/Fe] abundance, while
also illustrating the accuracy that MUSE could bring to this kind of
investigations.

\end{abstract}

%

%
\section{Introduction}
\label{sec:intro}

When it comes to painting a comprehensive picture for the formation and
evolution of galaxies, one of the key ingredients to consider is the
mass distribution with which stars initially form out of their giant
cradles of cold, molecular gas. 
For instance, measuring such an initial stellar-mass function (IMF) in
the optical regions of galaxies allows to weigh the relative fraction
of stellar and dark matter, which helps in understanding how dark and
baryonic matter interact \citep[e.g.,][]{Aug10,Son12}.
Constraining the IMF of passively-evolving stellar systems makes it
also possible to reconstruct their luminosity evolution and thus
correctly interpret the cosmic evolution of the most massive galaxies
in the Universe.
Finally - by providing the ratio of high-to-low mass stars - the form
of the IMF offers an handle on the amount of energetic feedback that
star formation can re-inject in the interstellar medium and thus
contribute to regulate the formation of galaxies
\citep[e.g.][]{Pio11,Fon17}.
%

Despite its importance, an exhaustive theory for the origin of the IMF
it still lacking, in particular for environments other than the disk
of the Milky Way or its globular clusters.
In fact, whereas in our immediate galactic neighbourhood there is
little evidence for variations in the IMF \citep[see, e.g.,][]{Bas10},
in the last few years a number of studies have claimed a steepening at
the low-mass end of the IMF in the most massive galaxies, either from
the analysis of integrated spectra
\citep[e.g.,][]{vDo10,Smi12,Con12b,LaB13,Spi14,Spi15a,McD14} ,
gravitational lensing \citep[e.g.,][]{Tre10,Spi12,Spi15b,Lei16}, or
modelling of the stellar kinematics \citep[e.g.,][]{ThoJ11,Cap12}.

Theoretically, one way of obtaining such a bottom-heavy IMF is to
consider molecular clouds characterised by supersonic turbulence,
which would lead to more fragmentation and to a lower characteristic
mass of the IMF \citep[e.g.,][]{Pad02,Hop13,Cha14}.
In this picture, such a supersonic turbulence would be driven by a
large rate of supernovae (SN) explosions sustained in turn by a high
rate of star formation, which would be consistent with the current
view whereby most of the stars in massive early-type galaxies (ETGs)
formed early-on during very intense star-formation episodes
\citep[e.g.,][]{ThoD05}.
Alternatively, a bottom-heavy IMF could result from a high pressure or
metallicity in the interstellar medium (ISM), within models where the
bulk of the IMF is set by radiation feedback \citep[e.g.,][]{Kru11}.
Either way, a time-dependent scenario seems to be required in order to
match the stellar metallicity and $\alpha$-element abundance of
massive ETGs, where an early and intense star-bursting stage actually
characterised by a top-heavy IMF precedes a relatively more prolonged
star formation phase with a bottom-heavy IMF \citep{Wei13,Fer15}.

The early findings that the variation towards a bottom-heavy IMF
correlates with the overabundance in $\alpha$-elements observed in the
old stars of massive ETGs \citep{Con12b,Smi12} would support the
previous turbulence-driven picture, as indeed such a characteristic
[$\alpha$/Fe] abundance pattern requires short star-formation
timescales that go in the direction of driving high star-formation
rates.
On the other hand, other works point toward a strong correlation with
stellar velocity dispersion $\sigma$ for the variation of the IMF
\citep{Spi12,Spi14,Cap12}, which is more difficult to interpret, in
particular since $\sigma$ is known to correlate with other global
properties of ETGs, such as mass, luminosity, metallicity [Z/H] and
[$\alpha$/Fe] abundance.
It is also possible that such a trend with $\sigma$ indicates an even
closer tie with the local depth of the potential well. This would
indeed be similar to the case of the global correlation between the
Mgb index and $\sigma$ \citep{Dav93}, when Mgb turned out - thanks to
integral-field data - to follow more closely the escape velocity
$V_{\rm esc}$, both locally and in a similar way across different
galaxies \citep{Sco09}.
In fact, the recent finding of a tight correlation between total
metallicity and dwarf-to-giant ratio \citep{Mar15b} could also be
consistent with a close link between this ratio and the local depth of
the potential well. The Mgb - $V_{\rm esc}$ correlation is indeed
predominantly driven by the variation of [Z/H] with $V_{\rm esc}$, in
particular in old stellar populations (Scott et al. 2013), so that the
trend between metallicity and low-mass end IMF slope may indirectly
result from correlations between [Z/H] and $V_{\rm esc}$ and $V_{\rm
  esc}$ and IMF slope.

At this point it is important to realise that previous IMF findings
are for the most part based on large-aperture measurement or global
dynamical models where all information internal to those systems is
lost.
An handful resolved studies have appeared as of today
\citep{Mar15a,LaB16,Men16,vDo17,LaB17}, and whereas many report the
presence of IMF gradients indicating more dwarf-rich stellar
populations towards the center other studies find little or no
evidence for radial IMF variations (see, e.g., \citealp{Mar15c} for
the special case of the compact ETG NGC1277 and the work of
\citealp{Alt17}, who also report no evidence for a bottom-heavy IMF
throughout their sample), stressing the importance of also accounting
for radial gradients in the abundance of elements entering the
IMF-sensitive absorption-line features (such the abundance of Sodium
entering the NaI feature at 0.81nm, \citealp{Zie15,McC16}).

Stepping in this direction, in this paper we present high-quality
integral-field spectroscopic observations of the giant elliptical
NGC~4486 (M87) obtained with the Multi-Unit Spectroscopic Explorer
\citep[MUSE,][]{Bac10} on the Very Large Telescope during the first
Science Verification observing run for this instrument.
Using the unique wavelength coverage and sensitivity of MUSE, we
extract high-quality aperture spectra and combine the analysis of
dwarf-sensitive features such as those defined in \citep{LaB13} and
\citet{Spi14} with constraints on the stellar age, metallicity and
      [$\alpha$/Fe], as well as individual abundance ratios, based on
      standard optical absorption-line indices \citep{Tra98,Cen01}.
This allows us to set firm constraints on the IMF as a function of
radius, which we compare to radial variations for the stellar age,
metallicity, [$\alpha$/Fe] abundance and stellar velocity
dispersion. 
In the process we also explore the need for a radial variation of the
[Na/Fe] abundance, which could provide insights on the yields of
odd-numbered elements in high-metallicity and alpha-enhanced stellar
populations, such as those resulting from the intense star-formation
episodes in central regions of ETGs \citep{Arr10}.
Constraining the [Na/Fe] abundance is also key to using the NaI
features at 0.81nm and at 1.14nm in order to constrain the IMF
low-mass end slope \citep{Zie15, Smi15, LaB17}.

This paper is organised as follows. 
In \S~\ref{sec:data} we briefly describe the MUSE science verification
observations and the reduction of the MUSE data. In
\S~\ref{sec:analysis} we introduce the annular and Voronoi-binned
aperture spectra that we use in our analysis and describe how we
measured in them the strength of standard and IMF-sensisive
asborption-line features, which includes also measuring in these
spectra the stellar kinematic broadening and the contamination from
nebular emission. Based on these measurements, in \S~\ref{sec:results}
we will then present evidence for a radial variation in the slope of
the IMF in M87 and our best constraints for a radial change in the
[Na/Fe] abundance, discussing in particular how such an IMF gradient
compares with the local variation of the stellar velocity dispersion
and other similar trends in the literature. Finally, in
\S~\ref{sec:conclusions} we draw our conclusions.

\section{MUSE Observations and Data Reduction}
\label{sec:data}

NGC~4486 (M87) was observed with MUSE, on June 28 2014, during the
Science Verification run for this instrument. In this paper we
consider only the central MUSE observations from programme 60.A-9312
(PI Sarzi) with two 1800s-long exposures, slightly offset and at a
right angle from each other to optimise the treatment of cosmic rays
and bad pixels.  Each of these on-source exposures were then followed
by a 900s-long exposure in the outskirts of M87 and by an even further
sky pointing, approximately 15\arcmin\ away from the center.

The final reduction of the science verification data was performed
using the v1.6 version of the standard MUSE pipeline, which works on
pixel tables. This includes creation of the master bias, flat field
and arc calibration solutions based on the standard calibrations
observed the same night.
These were used to prepare the science exposures for post processing
and trace of the spectra on the CCD, where we used the geometric and
astrometric solutions from MUSE commissioning runs to transform the
location of the spectra into the focal plane spatial coordinates. On
the night of the observations of our data, a standard star (CD-32
9927) was observed, which did not return a well suited response
function. Therefore, we used a better-suited white dwarf (EG\,274)
from the previous night to create a response function, and applied it
to both target exposures.

Sky emission was removed on each exposure separately, using the offset
sky field observations. The procedure consists in two steps, where we
first created a representative sky spectrum consisting of an estimate
of the sky continuum and sky lines. These were then adapted to and
removed from each of the science exposures, incorporating the
information of the instrumental line spread function for each of the
MUSE channels (IFUs). We noticed that better results were achieved if
weak OH absorption bands close to wavelength of the laboratory
H$\beta$ line were excluded from the creation of the sky spectrum, and
therefore we used a list of skylines starting from 5197.9 \AA.
Sky-subtracted pixels tables were then merged into a combined table
and finally used to create a standard data cube. This encompasses
approximately 1\arcmin\ square sampled in $0\farcs2\times0\farcs2$
spaxels and covers the optical domain from about 4749\AA\ to
9348\AA\ with a spectral resolution of $\sim$60\kms at
5500\AA\ \citep[see, e.g.,][]{Kra15}

One other major difference from the standard pipeline reduction was in
the treatment of the telluric absorptions. When running the final
processing of the MUSE pixel tables (task {\tt muse\_scipost}, we did
not pass a telluric correction function, as we did not have a good
template to prepare it. Instead, all spectra analyzed in this work
were further corrected for atmospheric telluric absorption using the
software {\sc molecfit} \citep{Sme15,Kau15}. For a given observed
spectrum, {\sc molecfit} computes a theoretical absorption model by
fitting regions of the spectrum affected by prominent telluric lines,
based on a radiative transfer code, and an atmospheric molecular line
database. The correction was performed by running molecfit on each of
the annular-aperture and Voronoi-binned spectra described in
\S\ref{subsec:annular_aperture} and \S\ref{subsec:SN300_bins},
obtaining subpercent-level fitting residuals in all spectral regions
relevant to our study and in particular around those IMF-sensitive
features that are most affected by telluric absorption.
To illustrate the accuracy of the {\sc molecfit} correction, in
Fig.~\ref{fig:one} we show our annular aperture spectra before and
after applying this procedure in the spectral region around both the
TiO2 and NaI IMF-sensitive features, at 6230\AA\ and 8190\AA,
respectively.

\placefigone

We note that even without considering spatial binning the quality of
the MUSE data is already quite good. At the edge of the MUSE
field-of-view (i.e., at a radius of 30\arcsec\ and for a V-band
surface-brightness of 19.7 magnitudes), one hour of observing time
allowed us to reach a median value for the signal-to-noise ratio
($S/N$, based on formal uncertainties on the spectra) per spaxel and
per wavelength resolution element of 15 and 25 at 5100\AA\ and
6400\AA, respectively.

\placefigtwo
\placefigthree

\section{Spectral Analysis}
\label{sec:analysis}

Detecting a variation in the number of low-mass stars in external
galaxies requires spectra of excellent quality. The signature of dwarf
stars can indeed be discerned only once high-precision measurements
for the strength of several absorption-line features can be secured
and analysed together. In the following we therefore combined up to
several hundreds or thousands of MUSE spectra in order to probe with
high-quality spectra the radial variation of IMF-sensitive features.

\subsection{Annular Aperture Spectra}
\label{subsec:annular_aperture}

In order to trace radial gradients for the IMF in M87, we started by
exploring whether the spherical symmetry of this galaxy could be
exploited and co-added the MUSE spectra in nuclear regions and in
eight nearly-circular annular apertures. This was done without
applying any redshift correction since M87 displays very little
stellar rotation (up to $\sim$20 \kms) compared to large values for
the stellar velocity dispersion (larger than $\sim$250 \kms\ within
the MUSE field of view; \citealp{Ems14}, hereafter EKS14).
Fig.~\ref{fig:two} shows the layout of these aperture on the MUSE
reconstructed image of M87, where regions affected by the non-thermal
emission from the jet have been excluded. Non-thermal emission from
the central active nucleus affects also the central aperture spectrum,
which will thus serve only for illustrative purposes.
We then proceeded to match these spectra with the {\sc pPXF\/} and
{\sc GandALF\/} programs \citep{Cap04,Sar06} in order to measure the
kinematic broadening of the stellar absorbtion features and check for
the presence of ionised-gas emission, both of which are elements to
consider when measuring the strength of stellar absorption lines.
In order to achieve the best possible match of the stellar continuum,
we used the entire MILES stellar library \citep{San06,Fal11} during
both the {\sc pPXF\/} and {\sc GandALF\/} fits, while also adopting a
10$^{\rm th}$-order polynomial correction, additive and multiplicative
for {\sc pPXF\/} and {\sc GandALF\/}, respectively.
Consistently with EKS14, the {\sc pPXF\/} fit was restricted to the
5050 -- 6000\AA\ wavelength range, whereas the present {\sc GandALF\/}
fits extended from 4800\AA\ to 6850\AA.

\placetabone

The basic properties of these nine aperture spectra are listed in
Tab.~\ref{tab:one}, whereas Fig.~\ref{fig:three} shows our best {\sc
  GandALF\/} fit to them.
Our models appear to match quite well such spectra, which are all of
very good quality as expected considering that our apertures encompass
several hundreds to thousands single MUSE spectra. Based on the
propagation of the statistical uncertainties on the flux density
values of the MUSE spectra, we should indeed be able to reach very
high formal values for the $S/N$ ratio, between 1500 and 2500, in our
annular aperture spectra.
Yet, the level of fluctuations in the residuals of our best {\sc
  GandALF\/} fit greatly exceeds such purely statistical errors, with
residual-noise levels that lead to signal-to-residual-noise values
($S/rN$) around just 240. The quality of {\sc pPXF\/} is considerably
higher, thanks to the use of additive polynomials and of a smaller
fitting range, but again this leads to $S/rN\sim 340$ that are still
quite short of the formal $S/N$ expectations.
Such a discrepancy cannot simply be ascribed to intrinsic limitations
in the modelling of the stellar continuum. For instance, in the case
of Sloan Digital Sky Survey (SDSS) data, Johansson et al. (2014) could
reach much higher $S/rN$ values (around $\sim 500$) when using the
MILES library to fit high $S/N$ stacked spectra of galaxies. 
Instead, the discrepancy between formal $S/N$ and observed $S/rN$
values relates most likely to small instrumental differences between
the 24 integral-field units in MUSE or to correlations between
adjacent spaxels introduced by interpolation processes in the MUSE
data reduction \citep[see also, e.g.,][for the case of CALIFA
  data]{Gar15}.

\subsection{Voronoi-Binned Spectra}
\label{subsec:SN300_bins}

Given the previous shortcomings in co-adding the MUSE spectra over such
large spatial regions as our annular aperture we decided to measure
the strength of absorption-line features also in Voronoi-binned
spectra similar to those that served to map the stellar kinematics of
M87 in EKS14.
For this we used the algorithm of \citep{Cap03} to spatially bin our
MUSE spectra to a target $S/N=300$ as done in EKS14 and then proceeded
to fit those spectra with {\sc pPXF\/} and {\sc GandALF\/} as we did
for our annular templates except as stellar templates were
concerned. In fitting the Voronoi-binned spectra we indeed use the
optimal templates that result from the best combinations of the MILES
stellar templates obtained during our {\sc pPXF\/} and {\sc GandALF\/}
fits to the annular aperture spectra, using such optimal templates for
the corresponding {\sc pPXF\/} and {\sc GandALF\/} fits to our binned
spectra.
The quality of these fits is also quite high, with average $S/rN$
values of around 230 and 130 for the {\sc pPXF\/} and {\sc GandALF\/}
fits, respectively. These $S/rN$ do not differ much from those
measured in our annular apertures, which again is indicative of
systematic variations across the MUSE field that do not average out
when spectra are combined in very large apertures, in addition to
spatial correlations introduced during the data reduction and inherent
limitations in our stellar continuum modelling.

Since the $S/rN$ values obtained in the annular-aperture spectra do
not significantly exceed the corresponding $S/rN$ values in the
Voronoi-binned spectra we do not expect the former spectra to lead to
particularly tighter constraints on the slope of the IMF. On the other
hand, the finer spatial sampling of the Voronoi-binned spectra allows
us to better investigate radial IMF trends. Furthermore, given the
circular symmetry of M87, the Voronoi-binned spectra effectively
provide independent radial index measurements that could be used to
derive realistic estimates for the error in our line-strength
measurements and thus, eventually, for low-mass end slope of the IMF
at any given radius.

For these reasons we will use the Voronoi-binned spectra to carry out
our stellar-population analysis (\S~4), and use the annular-aperture
spectra only for illustrative purposes in the remainder of this
section.

\subsection{Line-Strength Measurements}
\label{subsec:LS_measurements}

\placefigfour
\placetabtwo

For all extracted spectra, we measured the strength of several
IMF-sensitive features such as aTiO, TiO1 and TiO2 in the optical, as
well as the strength of Calcium triplet (using the Ca1, Ca2, and Ca3
indices) and of the NaI$\lambda\lambda 8183,8195$ absorption lines in
the near-Infrared.
Moreover, we measured line-strengths for the classic Mgb, Fe5015,
Fe5270, Fe5335 Lick indices in order to constrain the metallicity and
[Mg/Fe] abundance ratio, as well as the strength of NaD$\lambda\lambda
5890,5896$ doublet, which is key to gauging [Na/Fe] abundance ratios.
We also measured \hbo, a modified \hb\ index designed to minimize the
age-metallicity degeneracy \citep{Cer09}. The Mgb, Fe5270, Fe5335
indices were combined into the total metallicity indicator \mgfep$\rm
=[Mgb \times (0.72 \times Fe5270+0.28 \times Fe5335)]^{1/2}$ defined
in \citet{ThoD03}, which we used used in our analysis together with
the combined iron index \fem$=(Fe5270+Fe5335)/2$.
For aTiO, we followed the index definition of \citet{Spi14}, but took
care to interpolate the spectra in the rest-frame wavelength region
between $\lambda=5542.4$ and $5559.6$~$\AA$ in order to exclude the
residuals from the subtraction of the bright \OI\ 5577 sky emission
line. Notice that when comparing observations to predictions of
stellar population models (Sec.~4.2), the same interpolation is also
applied to the models before measuring line-strength indices.
For TiO1, we followed the Lick definition, while for both TiO2 and
NaI, we used the definitions of \citet{LaB13} where in particular for
TiO2 the red pseudo-continuum passband was slightly modified to avoid
sky-subtraction residuals from the prominent \OI\ 6300 sky line.

The definition of our main IMF-sensitive absorption features is
summarized in Tab~\ref{tab:two}, with their position indicated in
Figs.~\ref{fig:one} and \ref{fig:three}.  Notice that in the present
work we analyze all three Calcium-triplet lines seperately, rather
than combining them up into a single line strength. Since the three Ca
lines respond differently to the [Ca/Fe] abundance and the IMF slope
this approach allows us to disentangle the two effects, circumnventing
the need for bluer Ca spectral features such as Ca4227 that do not
fall within the MUSE spectral range (see \citealp{Men16}, for
details).

All index measurements were carried out without performing any
smoothing of the observed spectra. Instead, during the stellar
population analysis, for each spectrum we convolved the stellar
population models by the stellar velocity dispersion measured during
the {\sc pPXF\/} fit, which in the case of Voronoi bins yields a
stellar kinematics that is entirely consistent with the one published
in EKS14. 
This approach extracts the maximum amount of information from the
data, and avoids possible contamination of the relevant
absorption-line features from the residuals of the subtraction of
nearby sky lines when instead all observed spectra are smoothed to the
same stellar velocity dispersion.  We also did not place our indices
on the Lick system, as we rely on stellar population models based on
flux calibrated stellar spectra (i.e., the MILES stellar library).

Finally, we flagged and excluded from the remainder of our analysis
all Voronoi-binned spectra showing the presence of emission lines
affecting our absorption line-strength indices or with a significant
contribution from the non-thermal continuum associated to either the
jet or the active nucleus of M87. 
As regards nebular emission, despite the well-known presence of
ionised-gas emission in M87 \citep[e.g.,][]{Mac96,Sar06} we note that
none of our chosen IMF-sensitive indices (Tab.~\ref{tab:two}) could be
affected by it. 
On the other hand accounting for the presence of nebular emission
lines is crucial for our age, metallicity and [$\alpha$/Fe] abundance
estimates, in particular due to H$\beta$, [{\sc
    O$\,$iii}]$\lambda\lambda4959,5007$ and [{\sc
    N$\,$i}]$\lambda\lambda5197,5200$ emission entering the \hbo,
Fe5015 and Mgb indices, respectively.
Based on our {\sc GandALF} fit results we decided to take a rather
conservative approach and flagged Voronoi-binned potentially affected by
H$\beta$, [{\sc O$\,$iii}] and [{\sc N$\,$i}] emission only when the
much stronger [{\sc N$\,$ii}]$\lambda\lambda6548,6584$ lines were
comfortably detected.
Such a conservative approach proved more reliable compared to the
stardard procedure for judging the detection based on the value for
the line amplitude to residual-noise level ratio $A/rN$ \citep{Sar06},
and is further justified considering the old age of the stellar
population of M87 and thus for the need to deal with even very small
amounts of emission-line infill, in particular for the \hbo\ index.
As for the non-thermal continuum, we excluded bins inside a radius of
3\arcsec\ and along the jet direction, that is, within 7$^{\circ}$ of
its direction at a PA=-69$^{\circ}$ (as shown in Fig.~\ref{fig:two}).
The adverse impact of both nebular emission and the jet/AGN
non-thermal continuum on our \hbo, Fe5015 and Mgb indices can be
appreciated in Fig.~\ref{fig:four}, which otherwise display rather
tight and smooth radial gradients.

\placefigfive
\placefigsix

\placefigseven
\subsection{Line-Strength Gradients}
\label{subsec:LS_gradients}

Focusing on all but two of our key IMF-sensitive features and on the
NaD index (which is also somehow sensitive to the IMF),
Fig.~\ref{fig:five} allows already to appreciate the radial variation
for their strength across our annular aperture spectra.
This is then shown more quantitatively by Fig.~\ref{fig:six} where
the line-strength values are plotted against radius for the
Voronoi-binned spectra, excluding at this point regions affected by
nebular emission or a jet/AGN non-thermal continuum as discussed in
\S3.3.
Fig.~\ref{fig:six} shows that all our IMF-sensitive absorption-line
indices display clear radial gradients, which also allows to identify
Voronoi-bins to be further excluded from our analysis (mostly at the
edge of the MUSE field of view), whenever the values for the
line-strength indices are found to be significant (at a 3$\sigma$
level) outliers from a 3rd-order polynomial fit to the observed radial
trend.
Fig.~\ref{fig:six} also shows that the index values determined in the
annular aperture spectra agree fairly well with the median values
computed from the Voronoi bins at the same radial intervals, well
within the scatter of these last measurements that will also serve as
errors for our stellar-population analysis.

The radial gradients shown in Figs.~\ref{fig:five} and \ref{fig:six}
may already indicate a variation in both the IMF slope and the [Na/Fe]
abundance, although gradients in stellar metallicity or the abundance
of $\alpha$-elements ([$\alpha$/Fe]) could also contribute to them.
For instance, \citet{Spi14} shows that the strength of the aTiO, TiO1
and TiO2 indices increases not only with the IMF slope but also with
[$\alpha$/Fe]. Similarly, the NaD index responds very strongly to
metallicity in addition to the abundance of Sodium, whereas classical
metallicity- or [$\alpha$/Fe]-sensitive indices such as Fe5270 and Mgb
are far less sensitive to the IMF slope.
In fact, no single IMF-sensitive index allow to directly estimate the
IMF slope, and in order to constrain the low-mass end of the IMF this
parameter has to be optimised together with at least the stellar
metallicity and [$\alpha$/Fe] abundance (for an old galaxy such as
M87) while matching the strength of several spectral features that
together carry enough information on all these stellar population
properties, as detailed below.

\section{Stellar Population Analysis}

\subsection{Stellar population models}

To constrain the stellar population content of M87 we compared the
observed strength of the absorption-line indices in our MUSE spectra
to predictions from the MIUSCAT stellar population models
\citep{Vaz12,Ric12}.
More specifically, for each Voronnoi-binned spectrum of M87, we
compared observed and model absorption-line strengths derived after
smoothing the MIUSCAT models to the observed values for the stellar
velocity dispersion of the given spectrum, as obtained from {\sc
  pPXF\/} (see Sec.~\ref{subsec:LS_measurements}).

The MIUSCAT models cover the spectral range between 3465 and
9469\AA\ at a nominal resolution of $2.51$\,\AA\ \citep[FWHM][]{Fal11}
and provide state-of-art Simple Stellar Population (SSPs) predictions
based on the empirical stellar libraries, namely the MILES library in
the optical range ($\lambda\lambda 3525-7500$\,\AA; \citealp{San06})
and the CaT library in the Near-Infrared ($\lambda\lambda
8350-9020$\,\AA; \citealp{Cen01}). The Indo-US library \citep{Val04} is
then used to fill the gap between MILES and CaT.
The models rely on solar-scaled isochrones (either Padova00 or BaSTi;
from \citealp{Gir00} and \citealp{Pie04}, respectively), and are based
on stellar spectra following the abundance pattern of our Galaxy,
i.e. approximately solar-scaled at Solar metallicity and
$\alpha$-enhanced at metallicities below Solar \citep{Ben14}.
For comparison with our previous works \citep[e.g.,][]{LaB13,Spi14}, we
base our analysis on MIUSCAT models with Padova isochrones. The
MIUSCAT SSPs cover a wide range of ages, from $0.06$ to $17.78$~Gyr,
and seven metallicity bins, i.e.  \zh = -2.32, -1.71, -1.31, -0.71,
-0.4, 0.0 and +0.22. SSPs models are provided also for several IMFs
either unimodal (single power-law) or ``bimodal'' (low-mass tapered)
in flavour, which are both characterized by their slope, $\Gamma$
(unimodal) and $\rm \Gamma_b$ (bimodal), as a single free parameter
\citep[see, e.g.,][]{Vaz96,Vaz03}.
The ``bimodal'' IMFs are smoothly tapered off below a characteristic
``turnover'' mass of $0.6$\,M$_\odot$.
For $\rm \Gamma_b \sim 1.3$, the low-mass tapered IMF gives a good
representation of the Kroupa IMF, while for $\Gamma\sim 1.3$ the
unimodal IMF coincides with the Salpeter distribution.  The lower and
upper mass-cutoff of the IMFs are set to $0.1$ and $100$\,M$_\odot$,
respectively \citep[see also Fig.~3 of][for an illustration of these
  two IMF parametrisations]{LaB13}.
Notice that low-mass tapered IMFs have been shown to provide a better
description of the optical and NIR spectral features in ETGs
\citep{LaB16}, and of their mass-to-light ratios
\citep{Lyu16}. Therefore, we base the present analysis on MIUSCAT SSPs
with a low-mass tapered distribution.

\subsection{Fitting line-strengths}

As a first step, we proceeded to estimate the stellar age, metallicity
and the [$\alpha$/Fe] abundance ratio by applying the same approach as
in \citet[][hereafter LB13]{LaB13}. We start by estimating the stellar
age and metallicity in each of our Voronoi bins from the
\hbo\--\mgfep\ diagram, minimizing the rms of observed and model
line-strengths over a grid of SSP model predictions, interpolated over
a fine grid with steps of 0.05 Gyr in age and $0.006$ dex in
metallicity.
Then, at fixed age, we compute the solar-scaled proxy for
[$\alpha$/Fe], measured as the difference between metallicity
estimates from the Mgb and \fem\ line-strenght indices. Such a proxy
is calibrated onto [$\alpha$/Fe] with the aid of \citet{ThoD11}
stellar population models, as detailed in LB13, resulting into an
excellent accuracy (rms) of about 0.025 dex in [$\alpha$/Fe].
Notice that while the [$\alpha$/Fe] could be also estimated relying on
$\alpha$--enhanced stellar population models \citep[e.g.,][]{Vaz15},
such models do (partly) rely on theoretical atmosphere predictions,
while the proxy approach is entirely based on the empirical (MIUSCAT)
models. Moreover, the LB13 method for deriving [$\alpha$/Fe] is
independent of the adopted IMF (see \citealp{LaB16}, for more
details).
Fig.~\ref{fig:seven} shows the \hbo\--\mgfep\ diagram for the
Voronoi-binned MUSE spectra on which our analysis is based, which
excludes bins affected by nebular emission, a non-thermal continuum
associated to either the jet or AGN of M87 (see Fig.~\ref{fig:four}),
or where systematic effects (mostly near the edge of the MUSE field)
led to unreliable line-strength measurements (see Fig.~.\ref{fig:six}).
The \hbo\--\mgfep\ diagram of Fig.~\ref{fig:seven} shows that all
spectra of M87 are consistent with a very old age and a radial trend
that indicates the characteristic negative metallicity gradient
observed in ETGs, consistent also with a preliminary analysis of our
MUSE spectra with the STARLIGHT spectral fitting code of \cite{Cid05}.
In the following we will therefore considered only MIUSCAT models with
an old age of 14 Gyr, although we verified that none of the main
conclusions of this work change if we adopt a different age value or
consider age as an extra fitting parameter while constraining the IMF
slope (see also, e.g., the case of \citealp{LaB16}).
In particular, we note that if the data in Fig.~\ref{fig:seven}
suggest the presence in M87 of a positive age gradient and slightly
younger central stellar populations (which unfortunately would be
difficult to assess given how the data fall systematically below the
solar-scaled grid at all radii), should this be the case we would
infer an even steeper IMF gradient than by assumung a constant old
age, thus reinforcing the conclusions of our analysis.

\placefigeight 

Having constrained the stellar age, metalliticy and - most important -
the [Mg/Fe] ratio in our Voronoi-binned spectra, we proceed to
estimate the slope of the low-mass tapered IMF, \gammab, following a
similar procedure to LB13 and \citet{LaB15}. Specifically, we minimize
the expression:
\begin{multline}
\chi^2({\rm [Z/H], \Gamma_\mathrm{b}, [X/{\rm Fe}]}) = \\ \sum_i
\left[ \frac{ E_{{\rm corr}, i} - E_{{\rm M},i} - \sum_{\rm X}
    \Delta_{i, {\rm X}} \cdot {\rm [X/{\rm Fe}]}}{\sigma_{i}}
  \right]^2
\label{eq:method}
\end{multline}
where $E_{{\rm corr,}i}$ are the observed absorption-line strengths
for a selected set of spectral features (e.g., IMF-sensitive features)
corrected to Solar-scale based on our [Mg/Fe] estimates, $E_{{\rm
    M},i}$ are model absorption-line strengths for the same features,
${\rm [X/Fe]}$ is the abundance ratio of different chemical elements
and \Dix\ is the sensitivity of the i-th absorption-line index to a
given elemental abundance, i.e., \Dix$=\delta(E_{{\rm
    M},i})/\delta({\rm [X/Fe]})$.  Finally $\sigma_i$ is the
uncertainty on $E_{{\rm corr,}i}$, obtained by adding in quadrature
the error on our $E_i$ (as derived from the azimuthal variation at
different radii, see \S~3.4) with the uncertainty on our correction to
a Solar scale.

\placefignine

As regards the latter, we bring the observed absorption-line strengths
$E_i$ to a Solar scale following LB13, that is, we derived the trend
of a given index with [Mg/Fe] from SDSS stacked spectra of early-type
galaxies at fixed stellar velocity dispersion and use our previous
estimate of [Mg/Fe] in our Voronoi-binned spectra to correct the value
of the considered index to its [Mg/Fe]$=0$ value.
For this reason, the abundance ratios obtained from
Eq.~\ref{eq:method} should be interpreted as residual abundances
rather than absolute estimates of [X/Fe].
The advantage of this approach, as opposed to relying entirely on
models with varying [X/Fe] (such as, e.g., those of \citealp{Con12a}),
is that one relies as much as possible on empirical trends rather than
on more uncertain theoretical model calculations.
As for the $\rm E_{{\rm M},i}$, these are computed with MIUSCAT models
of varying metallicity and IMF slope, while the \Dix\ values are
derived from the publicly available stellar population models of
Conroy \& van Dokkum, taking models with a Solar metallicity, old age
($t=13.5$~Gyr) and a Kroupa IMF.
Notice also that since the Conroy \& van Dokkum models are computed at
fixed Fe abundance rather than total metallicity [Z/H], when applying
the \Dix\ terms in Eq.~1 one is actually changing the metallicity
content of MIUSCAT models, which enters the $\rm E_{{\rm M},i}$
values. Therefore, when solving for the IMF slope through Eq.~1 we
need to allow the total metallicity be an extra fitting parameter,
rather than fixing it to the best-fitting value from the
\hbo\--\mgfep\ diagram.  Similarly, since the abundance of single
elements is optimised while we look for the best IMF slope, the
best-fitting values of \zh\ from Eq.~1 should not be interpreted as
``true'' estimates of \zh.
To summarise, although the free fitting parameters in
Eq.~\ref{eq:method} are total metallicity \zh, IMF slope \gammab\
and elemental abundances [X/Fe], in practice our procedure is designed
only to provide an estimate for the IMF slope.

In order to prove the robustness of our results, we have performed the
$\chi^2$ determination procedure using different combinations of
IMF-sensitive indices and including different sets of element
abundance [X/Fe]. As shown below, we find that our main result,
i.e. the presence of a radial IMF gradient in M87, is independent of
the method adopted.
Finally, as regards the uncertainties on our best-fitting parameters,
these were obtained by a bootstrapping approach, whereby the fitting
is repeated after shifting the observed-line strengths and the values
of \mgfe\ (used to correct indices to \mgfe$=0$, see above) according
to their uncertainties.

\placefigten

\section{Results}
\label{sec:results}

\subsection{IMF, metallicity and [$\alpha$/Fe] abundance gradients}
\label{sec:gradients}

We start discussing the results of our stellar population modelling by
showing in Fig.~\ref{fig:eight} the radial profiles for the
best-fitting slope \gammab\ for a low-mass tapered ``bimodal'' IMF, as
obtained by mimimising the expression given by Eq.~1 while using
different sets of absorption-line indices and while either holding or
varying the abundance elements.

As a first step, in panel a) of Fig.~\ref{fig:eight} we show that a
tight \gammab\ gradient is found even when simply using only the
optical IMF-sensitive indices TiO1, TiO2 and aTiO and while holding
the abundance of all elements at a Solar scale.
\gammab\ is also found to be steeper at all radii than $\rm \Gamma_b=
1.3$, which corresponds to the case of a Kroupa, Milky-Way like IMF.
Since a varying [Ti/Fe] abundance may contribute to the observed TiO1,
TiO2 and aTiO gradients (Fig.~\ref{fig:five}) that in turn lead us to
infer a varying IMF slope, in panel b) we then show what happens when
we let the [Ti/Fe] ratio free. During this step we also include the
Fe5015 index in our analysis since this index responds negatively to
the [Ti/Fe] abundance \citep{ThoD11,LaB15}, so that adding Fe5015 helps
in breaking the degeneracy between IMF slope and [Ti/Fe] abundance.
Additionally, in panel c) we exclude the aTiO index since this is a
rather broad absorption-line feature thus making the value of the aTiO
index particularly sensitive to the accuracy of our relative flux
calibration.
Either ways, panels b) and c) show very similar \gammab\ profiles to
the one corresponding to our simplest approach of panel a), which
gives us confidence both in our treatment of the [Ti/Fe] abundance and
in our aTiO measurements.

Next, in panel d) of Fig.~\ref{fig:eight} we vary the [O/Fe] abundance
in addition to the [Ti/Fe] abundance, while using the TiO1, TiO2 and
Fe5015 indices but still leaving out aTiO as in panel c).
In panel e) we then further include the IMF-sensitive Ca1, Ca2 and Ca3
indices while allowing also for a varying [Ca/Fe] abundance, while in
panel f) we add the popular but IMF-sensitive NaI index, for which we
also need to account for [Na/Fe] variations that in turn are better
constrained when including also the NaD index in Eq.~1.
While panel d) still shows an IMF gradient, the \gammab\ values now
decrease more gently with radius and are significantly lower (on
average by 0.4) compared to what observed in panel c) when only the
[Ti/Fe] abundance was varied, which is expected given that both TiO1
and TiO2 respond positively to the [O/Fe] abundance.
The \gammab\ profiles of panels e) and f) are then remarkably similar
to those of panel d), which shows that both Ca and Na features point
to a radial IMF gradient consistent with that inferred from TiO
features.
This consistency also confirms the quality of our correction for
telluric absorption (Fig.~\ref{fig:one}).

Finally, in panels g), h) and i) we show how by reintroducing the aTiO
index in our analysis brings back the \gammab\ values to levels
similar to those observed in the top panels of Fig.~\ref{fig:eight}
where the [O/Fe] abundance was not varied, which reflects the fact
that aTiO responds weakly or negatively to [O/Fe] and thus the
importance of including it to disentangle the effects of [O/Fe] and
IMF.
Although not shown here for the sake of clarity, we also checked the
robustness of our results on the TiO1 and TiO2 features, by excluding
these two indices altogether from the fitting procedure and while
removing also Fe5015 and fixing [Ti/Fe] and [O/Fe] to Solar
values. Also in this case we found a clear IMF gradient very much in
line with the previously discussed cases.

Overall, the upshot from Fig.~\ref{fig:eight} is that irrespective of
our choice for the set of IMF-sensitive indices to include in our
analysis and for the set of elements abundance that we decided to
vary, one always finds that in the central region of M87 probed by our
MUSE observations the slope \gammab\ for a low-mass tapered
``bimodal'' IMF steeply decreases with radius, corresponding to a
fraction of low-mass stars that remains above what found in the Milky
Way.

Even though the model presented in panel i) of Fig.~\ref{fig:eight}
could be regarded as our best and final model, given that it includes
the largest set of IMF-sensitive indices and of varying element
abundances, we prefer to combine the IMF results from our different
model approaches and thus present in Fig.~\ref{fig:nine} (left panel)
a more conservative view of the IMF gradient in M87, which is also
meant to capture systematic effects both in our data and in stellar
population models.
The scatter around such combined \gammab\ values is still relatively
small, however, with average 68\% confidence limits of 0.28. This also
does not largely exceed average formal \gammab\ errors on our single
model approaches, which range between $\sim 0.1$ and $\sim 0.2$ and
could be regarded as the limiting accuracy on IMF measurements that
may be achieved with MUSE.

Fig.~\ref{fig:nine} (middle and right panels) also shows our best
values for the stellar metallity [Z/H] and $\alpha$-element abundance
[$\alpha$/Fe], which both exhibit rather tight and negative gradients with
a scatter of just $\sim 0.02$ dex.
The observed metallicity gradient is in line with the typical decrease
of -0.28 dex per decade in radius observed in ETGs
\citep[e.g.,][]{Kun10} and both our central, luminosity-weighted [Z/H]
and [$\alpha$/Fe] values (see Fig.~\ref{fig:nine}) agree well the
SAURON measurements of \citet{McD15} inside one-eight of the effective
radius ($\rm R_e/8 = 10\farcs2$, taking $R_e = 81\farcs2$ as in
\citealp{Cap11}).
At larger radii, however, our [Z/H] gradient is somehow steeper than
the one found by McDermid et al., who also report little or no
evidence for a gradient in [$\alpha$/Fe] in M87 \citep[consistent with
the previous SAURON analysis in this object by][]{Kun10}.
As regard the discrepancy in [$\alpha$/Fe] between us and McDermid et
al., we have verified that accounting for such differences (e.g., by
artificially increasing [$\alpha$/Fe] in our model by 0.1 dex) would
lead only to negligible changes the \gammab\ values (e.g., by a few
percent) inferred at large radii.

Interestingly, the metallicity gradient appears to change slope and
flatten at around the same distance of 7\farcs5 where the surface
brightness profile of M87 departs from a single Sersic law and we
observe the onset of a central core, as generally found in the most
massive ETGs \citep{Fer06}.
Such a coincidence would be consistent with the idea that galactic
cores result from the scouring of the central stellar regions due to
the formation of a supermassive black hole binary during a merger
event \citep[e.g.,][]{Mil01}, as indeed one would expect to also find
a similar flattening of any pre-existing gradient for the stellar
population properties.
In fact, we note that both the IMF and [$\alpha$/Fe] gradient seems
also to flatten towards the center, although [Z/H] appears still to
follow more closely the surface brightness than \gammab\ or
[$\alpha$/Fe]. The values of the Spearman's rank coefficient $\rho$
for a correlation between the surface brightness and [Z/H], \gammab\
and [$\alpha$/Fe] are indeed 0.97, 0.87 and 0.89, respectively, and
are all highly significant. 

Given such similarities between the radial trends shown in
Fig.~\ref{fig:nine}, it is perhaps not surprising to find in Fig.~10
(top panel) that our MUSE measurements for M87 parallel rather well
the IMF--metallicity relation that \citet{Mar15b} derived using both
resolved (from \citealp{Mar15a} and from the CALIFA survey,
\citealp{San12}) and unresolved measurements (from SDSS data) in ETGs.
Even thought the agreement with the IMF--metallicity trend of
Mart\'{i}n-Navarro et al. supports their suggestion that the IMF shape
is more tighly connected to the stellar metallicity then to other
stellar population parameters or physical quantities such as stellar
velocity dispersion, we first of all note that in our case the
\gammab\ also follows [$\alpha$/Fe] fairly well (Fig. 10, middle
panel).
Furthermore, if there are reasons to consider the role of metallicity
in determining the local stellar mass spectrum (e.g., given that
conversely metal-poor systems show evidence for a top-heavy IMF,
\citealp{Mar12}) we note that in first place the metallicity gradient
of ETGs was likely set up by the local depth of the potential well
during the same very intense and short episode of star-formation that
led to enhanced [$\alpha$/Fe] ratios throughout the galaxy \citep[see,
  e.g.,][]{Pip10}.
This is indeed particularly evident when the stellar metallicity is
compared with the luminosity-weighted predictions for the escape
velocity $V_{\rm esc}$, either globally \citep{Dav93} or locally
within galaxies \citep{Sco09,Sco13}.\footnote{In this respect, we note
  that our \gammab\ values follow the $V_{\rm esc}$ values provided by
  the models of Scott et al. (2013) just as well as they follow our
  [Z/H] estimates (with a Spearman's rank coefficient $\rho$ = 0.87),
  even though a proper comparison between IMF slope and $V_{\rm esc}$
  would require the latter to be computed self-consistently on the
  basis of mass models accounting for a varying mass-to-light ratio.}
On the other hand, our MUSE measurements (Fig.~10, lower panel)
confirm the finding of Mart\'{i}n-Navarro et al. that the local
stellar velocity dispersion $\sigma$ does not trace well the radial
variation of the IMF in ETGs. In M87 this is particularly evident
since the $\sigma$ radial profile reaches a plateau at around 265
\kms\ from a radius of $\sim20\arcsec$ onward (Tab.~1, but see also
Fig.~3 from \citealp{Old17} where $\sim20\arcsec$ correspond to
$\sim1.5\rm\, kpc$), whereas the \gammab\ values continue to decrease
beyond this point (just like our [Z/H] and [$\alpha$/Fe] values do,
see also Fig.~\ref{fig:nine}).

\subsection{Mass-to-light ratio and mismatch parameter}
\label{sec:mlgradient}

In order to further compare our results with previous IMF studies,
including those based on dynamical constraints for the IMF, we
evaluate the mass-to-light ratio $M/L$ that corresponds to our best
stellar population results. Similarly, we also compute the so-called
mismatch parameter $\alpha_{\rm IMF}$ by taking the ratio of our best $M/L$
values to those that would correspond to a Milky-way like IMF while
holding to our best stellar age, metallicity and [$\alpha$/Fe]
estimates.
Fig.~\ref{fig:eleven} (top panel) shows the radial profile for our
best $M/L$ estimates in the SDSS r-band as well as the r-band
$(M/L)_{\rm Kr}$ values that would correspond to a Kroupa IMF, which
allows to evaluate the mismatch parameter $\alpha_{\rm IMF, Kr} =
(M/L)/(M/L)_{\rm Kr}$ as a function of radius (lower panel).

\placefigeleven
\placefigtwelve

The presence of an higher fraction of low-mass stars towards the
center of M87 has a strong impact on the projected $M/L$ values, more
than doubling $M/L$ from 40\arcsec\ to 4\arcsec and well exceeding
what would derive from changes only in stellar metallicity or
$\alpha$-element abundance.
Such a dramatic $M/L$ variation ought to be included in dynamical
models, in particular when considering that intrinsically, i.e. when
deprojected, the $M/L$ gradient in M87 is likely to be even steeper
and to extend to even higher $M/L$ values towards the center.
For instance, giving more weight to stars in the central regions of
M87 may help with the current tension between gas- and
stellar-dynamical measurements for the mass of the central
supermassive black hole, where the latter have historically exceeded
the former by nearly a factor two (see, e.g., \citealp{Old16} and
\citealp{Geb11} vs. \citealp{Wal13} and \citealp{Mac97}).

In the lower panel of Fig.~\ref{fig:eleven} we also compare our
$\alpha_{\rm IMF, Kr}$ radial profile to the mean trend found by
\citet{vDo17} for their sample of six ETGs and to the $\alpha_{\rm
  IMF, Kr}$ gradient measured by \citet{LaB16} in the galaxy XSG1,
both of which were derived through a stellar-population analysis.
The $\alpha_{\rm IMF, Kr}$ profile matches fairly well the
gradient observed in XSG1, which is also a very massive ETG like M87,
and remarkably agree with both the average trend reported by van
Dokkum et al. and the data of La Barbera et al. in that by
approximately 0.4--0.5 $R_e$ also in M87 we observe a nearly Milky-Way
like IMF.
As we approach the center our $\alpha_{\rm IMF, Kr}$ values
systematically fall below the line drawn by van Dokkum et al.,
however. This discrepancy may relate to their different
parametrization for the IMF whereby the low-mass end of the IMF is
allowed to vary (in two stellar-mass $m$ regimes between
$0.08M_{\odot}<m<0.5M_{\odot}$ and $0.5M_{\odot}<m<1.0 M_{\odot}$),
thus effectively resembling more like an unimodal form and hence
possibly leading to larger $M/L$ values.
Yet, we also note that there are significant differences between the
$\alpha_{\rm IMF, Kr}$ gradients observed in each of the van
Dokkum et al. targets. For instance, one of the two most massive
objects in their sample, NGC~1600, shows a flatter $\alpha_{\rm IMF,
  Kroupa}$ gradient reaching up to central values similar to what we
find in M87.

Turning now our attention to dynamical measurements of the IMF slope
based on direct estimates for the stellar $M/L$ ratio, in
Fig.~\ref{fig:twelve} we compare our radial profile for the mismatch
parameter with both the global estimates from the ATLAS$^{\rm 3D}$
survey \citep{Cap13} and the recent resolved measurements in M87 by
Oldham \& Auger (2017).
More specifically, in the top panel of Fig.~\ref{fig:twelve} we plot
the r-band mismatch parameter rescaled with respect to a Salpeter IMF
$\alpha_{\rm IMF, Sa}$ against stellar velocity dispersion $\sigma$,
in order compare our findings with both the corresponding $\alpha_{\rm
  IMF, Sa}$ and $\sigma$ global values (within 1$R_e$) for M87 and
to the trend between these same quantities acrossed different ETGs
(red square in Fig.~\ref{fig:twelve}).
Overall our $\alpha_{\rm IMF, Sa}$ value are consistent with the
global ATLAS$^{\rm 3D}$ data point, as the luminosity-weighted
integrated measurements inside $R_e/8$ and $R_e/2$ (orange points)
appear on course to fall not too far off from the ATLAS$^{\rm 3D}$
measurement within $R_e$.
However, as in the case of the IMF slope (Fig.~\ref{fig:ten}, lower
panel), the stellar velocity dispersion traces the mismatch parameter
only in the central regions. Furthermore $\alpha_{\rm IMF, Sa}$
increase with $\sigma$ faster than it is observed from integrated
measurements across different galaxies, further suggesting that
$\sigma$ is not an optimal tracer for IMF variations in ETGs.

In the lower panel of Fig.~\ref{fig:twelve} we show the radial profile
of our V-band mismatch parameter with respect to a Chabrier IMF
$\alpha_{\rm IMF, Ch}$ in order to compare this profile to the
corresponding gradient from the dynamical and stellar population
models for M87 of Oldham \& Auger, which is presently a unique study
in that it allows for a radially varying $M/L$ ratio in their
dynamical models.
Although the $\alpha_{\rm IMF, Ch}$ values of Oldham \& Auger lie
systematically below ours (on average by a factor 1.3), their
$\alpha_{\rm IMF, Ch}$ gradient parallels very well our $\alpha_{\rm
  IMF, Ch}$ profile.
Such an offset most likely originates from the requirement of the
Oldham \& Auger models that the dynamical $M/L$ values reach those
expected for a Chabrier IMF towards the outer parts of M87, which for
them correponds to regions 20\arcsec ($\sim$ 1.5 kpc) away from the
center where our stellar-population analysis still indicates an excess
of low-mass stars compared to the case of the Milky Way.
On the other hand, we also note that such a level of discrepancy falls
within the systematic factors involved in this kind of comparison
between spectral and dynamical constraints of the IMF. Indeed, for
several of their objects \citet{Lyu16} report spectral mismatch
parameter estimates that exceed the values of their dynamical
counterparts, by up to $\sim25\%$.
Despite the offset, the remarkable agreement in the slope of the two
independently-derived radial profiles for the mismatch parameter shown
in Fig.~\ref{fig:twelve} gives us more confidence on our model
assumption for a low-mass tapered ``bimodal'' IMF in M87.
High-quality data such as those presented here may indeed reveal
systematc IMF variations across different objects that could depend on
factors yet to be identified.

\subsection{The [Na/Fe] abundance}
\label{sec:sodium}

The behaviour of optical to near-infrared Sodium spectral features in
massive galaxies has recently received significant attention, not only
in relation to their sensitivity to the low-mass end of the IMF (in
particular for the NaI 0.81nm and 1.14nm features, see e.g.,
\citealp{Con12a,Spi14,Smi15,Zie15,McC16,LaB17}) but also in
connection to the detection of interstellar absorption and the
possible presence of cold-gas outflows \citep[around the Na~D lines at
  0.59nm, see, e.g.,][]{Jeo13,Par15,Sar16,Ned17}.
Yet, tackling the importance of a possible variation in [Na/Fe]
abundance is critical to properly interpreting the observed strength
of Sodium features and thus draw conclusions on the IMF slope or on the
properties of the ISM.
Super-solar [Na/Fe] abundance patterns are indeed to be expected in
massive ETGs as a result of their short star-formation histories
since, similar to the case of $\alpha$ elements, Sodium is mainly
produced in type II supernovae \citep{Tsu95}.
Furthermore, the presence of a bottom-heavy IMF in the low-mass
tapered ``bimodal'' form that we considered, may make the contribution
of intermediate-mass stars (between 3 and 8 $\rm M_{\odot}$) in their
Asymptotic Giant Branch (AGB) phase quite relevant
\citep[][]{Ven13,LaB17}.

In the MUSE spectral range the classical Na~D Lick index is the most
sensitive to changes in the [Na/Fe] abundance, responding also
strongly to stellar metallicity and only midly to the [$\alpha$/Fe]
abundance and the shape of low-mass end of the IMF.
This makes the NaD index ideal to trace the [Na/Fe] abundance as a
function of radius in M87, which is also known to contain little or no
dust \citep[e.g.][]{Bae10}.

To constrain [Na/Fe] abundance ratios, we rely on the Na-MILES stellar
population models recently implemented by \citet[][hereafter
  LB17]{LaB17}, which provide SSPs with varying [Na/Fe], up to
$\sim$1.2\,dex and over the same range of age, total metallicity, and
IMF slope as in the extended MILES models of \citet{Vaz15}
For each of our Voronoi-binned spectra in M87, we first correct the
Na~D line strength to a Solar $\rm [\alpha/Fe]=0$ scale, using the
following equation:

\begin{equation}
\rm
Na\,D_{corr} = Na\,D \cdot ( 1 + \alpha_{Na\,D} \cdot [\alpha/Fe] )^{-1}
\end{equation}

where

\begin{equation}
\rm \alpha_{Na\,D}=\frac{\delta(Na\,D)/Na\,D}{\delta[\alpha/Fe]}=-0.83
\end{equation}

is the relative response of the Na~D index to [$\alpha$/Fe]. The value
of ${\rm \alpha_{Na\,D}}$ was empirically determined in LB17 and the
fact that $\rm \alpha_{NaD}$ is negative shows that the NaD index
decreases with [$\alpha$/Fe] \citep[see, e.g.,][]{Con12a}.
Then, we derive the value of [Na/Fe] of the Na-MILES SSP model
prediction that fits better the $\rm NaD_{corr}$ values, by fixing
age, metallicity and IMF slope \gammab\ to the best-fitting values
that were previously derived in the given spectrum. 
In order to estimate the uncertainty on [Na/Fe], we reiterate such a
[Na/Fe] mimisation procedure by shifting the values of metallicity,
IMF slope, and [$\alpha$/Fe] according to their uncertainties.

\placefigNaD

Fig.~\ref{fig:NaD} shows the results of this procedure, which returns
[Na/Fe] abundance values well above Solar in the central regions of
M87 probed by our MUSE observations (that is, within $R_e/2$) and a
considerable negative [Na/Fe] gradient of around 0.15 dex per decade
in radius.
Such high [Na/Fe] values can be explained considering that Na is
primarily made in massive stars and ejected in Type II supernovae,
whereas Type Ia explosions produce very little of it
\citep{Tsu95}. Furthermore, intermediate-mass AGB stars ($\rm
3<M/M_{\odot}<8$) may have also returned significant amounts Na in a
short period of time (less than a 1 Gyr).
Thus, similar to the case of $\alpha$-elements, the
observed super-Solar [Na/Fe] values relate almost certainly to the
short timescales over which the stars in massive galaxies such as M87
are thought to have formed.

The inferred [Na/Fe] values indeed follow rather well the radial
[$\alpha$/Fe] profile shown in Fig.~\ref{fig:nine} although we note
that the [Na/Fe] gradient is somewhat steeper than the [$\alpha$/Fe]
gradient. To put it in another way, the [Na/$\alpha$] ratio, which
takes SN~Ia out of the picture and traces more directly the yields of
SN~II, increases towards the centre.
This behaviour could relate to metallicity, which would boost the
secondary production of Na in SN~II through neutron-capture processes
while having only a limited impact on the SN~II yields of
$\alpha$-elements \citep[e.g.][]{Kob06}. If the metallicity gradient
of M87 was also put in place early-on while the bulk of its stars
formed \citep{Pip10}, then it is possible that the early interstellar
medium was further enriched in Na towards the centre thanks to
metal-rich SN~II. In this respect, intermediate-mass AGB stars may
have played a role too, since their Na yields are also predicted to
increase with metallicity \citep{Ven13}.
On the other hand, we note that these results contrast with the
finding of \citet{LaB17} in the massive ETG XSG1, where the [Na/Fe]
radial profile inside $R_e/2$ is found to be similarly flat to that of
the [$\alpha$/Fe] ratio despite the presence of a significant
metallicity gradient. Hopefully the origin of such a discrepancy will
become clear as the number of ETGs with resolved gradient abundance of
Na and other elements increases.

To conclude this section we note that also the [Na/Fe] profile shows a
flattening towards the central regions, as observed for the \gammab,
[Z/H] and [$\alpha$/Fe] profiles plotted in Fig.~\ref{fig:nine}. This
further supports a scenario whereby the stellar-population properties
of M87 may have once shown steeper gradients towards the center,
taking the present shape during the formation of a supermassive black
hole binary and the scouring of central stellar regions that led to
the flattening of the surface brightness profile observed in this
massive galaxy.

%
\section{Conclusions}
\label{sec:conclusions}

Using MUSE integral-field spectroscopic data for the central regions
of the giant elliptical NGC~4486 we have investigated the low-mass end
of the stellar initial mass function in its stellar population. 
For this we have followed the approach of La Barbera et al. (2013,
2015) to match the predictions of state-of-the-art stellar population
models to the strength of various absorption-line features that are
sensitive to the fraction of dwarf stars, as observed after binning
our data to a target S/N=300 and in regions unaffected by non-thermal
or nebular emission.
Assuming a constant stellar age of 14 Gyr and a ``bimodal'', low-mass
tapered form for the IMF where the slope \gammab\ above a mass of 0.6
M$_{\odot}$ is free to vary, after solving for the best metallicity
[Z/H] and $\alpha$-elements abundance [$\alpha$/Fe] we find that:

\begin{itemize}

\item M87 displays a significant negative IMF gradient inside $R_e/2$,
  with ${\rm \Gamma_b > 1.3}$ at all radii indicative of an excess of
  low-mass stars compared to the case of the Milky Way, in particular
  towards the innermost regions.

\item The presence of such an IMF gradient is robust against the
  choice of the IMF-sensitive features that we include in our analysis
  (between aTiO, TiO1, TiO2, NaI and the Ca triplet features) and the
  set of elements abundance (among Ti, O, Na and C) that we decided to
  vary.

\item The IMF slope \gammab\ is found to closely follow the [Z/H]
  values, consistent with the findings of \citet{Mar15b} who suggested
  that metallicity traces IMF variations better than other
  stellar-population or physical properties. 
%

\item The stellar velocity dispersion $\sigma$, on the other hand,
  does not trace well the \gammab\ gradient in M87, further suggesting
  that locally, $\sigma$ is not a good tracer of IMF variations in
  ETGs \citep[see also][]{vDo17}.
%

\item The observed IMF gradient corresponds to a dramatic variation in
  the value of the projected stellar mass-to-light ratio $M/L$ in M87,
  which more than doubles between 4\arcsec\ and 40\arcsec, our probed
  radial range.

\item Accounting for variations due to changes in [Z/H] or
  [$\alpha$/Fe], this $M/L$ gradient corresponds to a radial profile
  for the mismatch parameter $\alpha$ which indicates that by $R_e/2$
  there is only a mild excess of low-mass stars compared to what found
  in the Milky-Way. Put together with the recent works of
  \citet{LaB16} and \citet{LaB17}, our results in M87 further suggests
  that IMF variations are a very central phenomenon in ETGs.

\item Remarkably, our $\alpha$ profile parallels very well the one
  derived by Oldham \& Auger (2017) in M87 based on $M/L$ measurements
  from dynamical models. This agreement further supports the choice
  for a bi-modal parametrization for the IMF over a uni-modal form,
  adding to similar comparisons between dynamical and spectral IMF
  measurements by \citet{LaB16} and \citet{Lyu16}.

\end{itemize}

In addition to constraining radial variations for the IMF in M87 we
also estimated the abundance of Sodium, not only because this is key
to interpreting the strength of near-IR IMF-sensitive features such as
NaI 0.81nm and NaI 1.14nm but also since tackling the [Na/Fe]
abundance could have bearings on ISM studies based on the Na~D
interstellar absorption.
Following the method of \citet{LaB17} we find that:

\begin{itemize}

\item M87 displays an Na abundance well above Solar inside $R_e/2$,
  reaching to [Na/Fe] values around 0.7 dex towards the centre, and a
  considerable negative [Na/Fe] gradient of 0.15 dex per decade in
  radius.

\item While such [Na/Fe] values are most likely the result of a short
  star-formation history and the prominent role of SN~II, as in the
  case of the observed super-Solar [$\alpha$/Fe] abundances, we note
  that the [Na/Fe] gradient is somehow steeper than the [$\alpha$/Fe]
  gradient. This may indicate an additional role for metallicity,
  which may have boosted the Na-yields in the central, metal-rich
  regions of M87.

\end{itemize}

Finally, thanks to the fine spatial sampling afforded by the quality
of our MUSE data we find that the radial profile of all our derived
stellar-population properties, i.e. the IMF slope \gammab, the
metallicity [Z/H] and the [$\alpha$/Fe] and [Na/Fe] abundance ratios,
appear to flatten towards the center of M87, at about the same
distance where the surface brightness profile breaks away and starts
to fall short from a single Sersic profile. 
Interestingly, such a coincidence would be consistent with the
currently accepted scenario for the formation of the central
surface-brightness cores of massive ETGs, following a merger event and
the ensuing hardening of a supermassive black hole binary.\\

Our work adds the case of the massive, central-cluster galaxy M87 to
the relatively few objects that as of today have radial constraints on
the low-mass end of the IMF, while also illustrating the accuracy that
MUSE integral-field observations can bring in this kind of studies
\citep[see also, e.g.,][]{Men16}.
High-quality data are indeed most welcome in this respect since,
nonewithstanding systematic differences in the stellar-population
modelling approaches, it already appears that ETGs may present rather
different IMF gradients, even when considering objects of similar mass
(e.g., the massive objects of van Dokkum et al. 2017 and La Barbera et
al. 2017). 
An exceptional data-quality will be also needed by future dynamical
modelling efforts which, similar to the work of Oldham \& Auger in
M87, may explore the presence of radial variations of the total
mass-to-light ratio in ETGs (see also, e.g., Davis \& McDermid 2017)
and thus provide a way to further test for stellar-population models.
In fact, assisted by adaptive optics, it will now be possible to use
MUSE to tackle IMF variations down to the nuclear regions of ETGs and
of the bulges of spiral galaxies. This may impact on measurements for
the mass of central supermassive black holes based on stellar dynamics
and help understanding the systematics differences between black-hole
mass measurements from different tracers of the gravitational
potential.

\section*{Acknowledgements}
We wish to thank the anonymous referee for providing valuable comments
to this work. MS is indebted to G. Cescutti, K. Gebhardt,
C. Kobayashi, S. Viaene and J. Walsh for their suggestions and to
L. Oldham and M. Auger for sharing with us their IMF results prior to
the acceptance of their work. MS also acknowledges the hospitality of
the European Southern Observatory where part of this work was carried
out. CS has received funding from the European Union's Horizon 2020
research and innovation programme under the Marie Sklodowska-Curie
actions grant agreement n. 664931. DK thanks P. Weilbacher for a
suggestion to use restricted sky list. This publication has made use
of code written by James R.~A. Davenport.

%

\label{lastpage}
\end{document}